\documentclass{aa}  

\usepackage{graphicx} 
\usepackage{xcolor}
\usepackage{comment}
\usepackage[normalem]{ulem}
\usepackage{hyperref}
\usepackage{siunitx}
\usepackage{txfonts}
\newcommand{\teff}{$T_{\mathrm{eff}}$}
\newcommand{\msun}{M$_\odot$}

\newcommand\omicron{o}
\newcommand{\AJ}[1]
{\textcolor{magenta}{#1}}

\newcommand{\referee}[1]{\textcolor{black}{#1}}
\newcommand{\logTc}{$\log\epsilon$\,(Tc)}
\renewcommand\omicron{o}
\newcommand{\TiO}{\mathrm{TiO}}
\newcommand{\ZrO}{\mathrm{ZrO}}

\newcommand{\totalM}{39}
\newcommand{\tcrichM}{18}
\newcommand{\tcpoorM}{21}

\begin{document} 
\titlerunning{Technetium-rich M Stars: Prime diagnostics of recent third dredge-up events on the Asymptotic Giant Branch}
\authorrunning{Shetye et al.}

\title{Technetium-rich M Stars: Prime diagnostics of recent third dredge-up events on the Asymptotic Giant Branch
}

   \author{Shreeya Shetye\inst{1} \and
          Sophie Van Eck\inst{2} \and 
          Alain Jorissen \inst{2} \and
          Ana Escorza \inst{3, 4}\and 
          Lionel Siess\inst{2}   \and
          Stephane Goriely\inst{2}   \and \\
          Hans Van Winckel \inst{1}\and
          Stefan Uttenthaler \inst{5}\and
          Nicolas Wijsen \inst{6}
          }
   \institute{ Instituut voor Sterrenkunde, KU Leuven, Celestijnenlaan 200D bus 2401, Leuven, 3001, Belgium\\
              \email{shreeya.shetye@kuleuven.be}
        \and
             Institute of Astronomy and Astrophysics (IAA), Université libre de Bruxelles (ULB),
              CP  226,  Boulevard  du  Triomphe,  B-1050  Bruxelles, Belgium
        \and
    Instituto de Astrofísica de Canarias, C. Vía Láctea, s/n, 38205 La Laguna, Santa Cruz de Tenerife, Spain.
    \and
    Universidad de La Laguna, Dpto. Astrofísica, Av. Astrofísico Francisco Sánchez, 38206 La Laguna, Santa Cruz de Tenerife, Spain.
         \and
             Institute of Applied Physics, TU Wien, Wiedner Hauptstra\ss e 8-10, 1040 Vienna, Austria
        \and
        Department of Mathematics, KU Leuven Kulak campus, Etienne Sabbelaan 53, 8500 Kortrijk, Belgium
             }

   \date{Received; accepted }

% \abstract{}{}{}{}{} 
% 5 {} token are mandatory
 
  \abstract
  % context heading (optional)
   {
   Technetium (Tc)-rich M-type stars have been known for over 45 years. However, the origin of Tc in these stars, particularly its detection without the concomitant detection of other s-process elements, typically produced during the asymptotic giant branch (AGB) s-process nucleosynthesis, remains poorly understood.
   Technetium was first identified in the spectra of S-type stars (which exhibit prominent ZrO bands) in 1952. The simultaneous enrichment of both Zr and Tc is well understood within the framework of s-process nucleosynthesis, which occurs during the AGB phase. However, despite being known for 45 years, Tc-rich M stars remain an enigma, as M-type stars are typically not enriched in heavy elements.}
   %leave it empty if necessary  
  % aims heading (mandatory)
   {This study aims at analyzing high-resolution spectra of a large sample of M-type stars to %(i) classify them as Tc-rich or Tc-poor \SVE{**I would remove this first item and out it in methods**}, 
   examine their spectral characteristics, and to compare their spectral properties with those of Tc-rich S-type stars in an attempt to understand the origin of their difference.}
   %Quantify the Tc abundance if possible (abundance analysis/Tc indices ratio). Locate the large sample in the HR diagram.  }
   {We define a robust classification scheme to assign M stars to the Tc-rich or Tc-poor class.
   We compute nucleosynthesis models to trace the evolution of Zr and Tc abundances across successive thermal pulses during the AGB phase. We further analyze 
   %measured and synthetic 
   spectral indices measuring the depth of the TiO and ZrO bands as well as the wavelength of the Tc blend
   on both synthetic and observed spectra.}
  % methods heading (mandatory)
   {The Tc lines in Tc-rich M stars are similar to those in S stars. However, Tc-rich M stars exhibit stronger TiO bands than S stars while displaying similarly strong ZrO bands. Spectral synthesis, together with location in the HR diagram and spatial properties suggest that Tc-rich M stars may have slightly lower metallicity and lower masses than Tc-rich S stars. 
   }
   {}
   

   \maketitle
%
%________________________________________________________________

\section{Introduction}

The unstable slow neutron capture nucleosynthesis (\textit{s}-process) element technetium (Tc) was first detected in the spectra of evolved stars by \citet{Merrill1952}.
Tc, along with other s-process elements, is synthesized via s-process nucleosynthesis during the thermally pulsing asymptotic giant branch (TP-AGB) phase. These elements are subsequently transported from the stellar interior to the surface through third dredge-up (TDU) episodes, where they become observable in stellar spectra. 
Since only $^{99}$Tc is produced in significant amounts by the s-process and has a half-life of 2×$10^5$ years, its detection in stellar photospheres serves as a clear indicator of ongoing s-process nucleosynthesis in TP-AGB stars.
This distinguishes such stars from extrinsic s-process-enriched systems, where the observed chemical enhancements result from past mass transfer from a former AGB companion in a binary system \citep{shetye2018}.
For a detailed review on s-process-enhanced stars we refer the reader to the recent review from \cite{Vaneck2022}.

Tc-rich M stars present a significant challenge in understanding TDU physics and AGB nucleosynthesis.
\citet{Little1979} first detected Tc lines in the spectra of several M-type stars. These stars are characterized by prominent TiO and VO bands, with only weak ZrO bands. As an AGB star evolves, recurrent TDU episodes enrich its surface with \textit{s}-process elements including zirconium (Zr), strengthening the ZrO bands relative to TiO. This marks the transition from spectral type M to S, where both TiO and ZrO bands are prominent.
As the star transports carbon (a product of He burning) to the surface through TDU, the star undergoes a chemical shift from oxygen-rich to carbon-rich composition, eventually becoming a carbon (C) star.
Thus, AGB stars are traditionally understood to follow the M → MS → S → SC → C spectral sequence.
The presence of Tc in M-type stars with weak or absent ZrO bands (indicating minimal s-process enrichment beyond Tc) suggests two possible scenarios. First, these stars may represent the earliest stages of the thermally-pulsing AGB phase, having experienced only one/a few thermal pulses and trace the onset of TDU. Alternatively, they could be sites of an alternative s-process mechanism in which Tc is produced without a corresponding enhancement of Zr \citep{Vanture1991}.

The discovery of a growing number of Tc-rich M stars has been made possible by the increasing availability of high-resolution spectra, which enable the detection and analysis of weak blue Tc~I lines. In contrast, earlier classifications of M and S stars relied primarily on TiO and ZrO molecular bands, which are prominent in low-resolution spectra. Hence the apparent scarcity of Tc-rich M stars does not necessarily mean that they are rare, but that they are challenging to detect.
In \cite{shetye2022} we conducted a pilot study on the Tc-rich M-type star S Her using high-resolution spectroscopy. In the present work, we expand upon that study by presenting a detailed spectroscopic analysis of \totalM~M-type stars to further investigate the spectral properties of this class of objects.
The sample selection and observational details are described in Sect.~\ref{sec:sample}. The detection of Tc within our sample is presented in Sect.~\ref{sec:tc}. 
The spectral characteristics of the sample stars are examined in Sects.~\ref{sec:Tclinefitting} and~\ref{sec:TiOZrO}. 
Subsequently, Sect.~\ref{sec:stellarparam} presents an analysis of the stellar parameters, metallicities, and positions in the Hertzsprung–Russell (HR) diagram for a subsample of stars.
In Sect.~\ref{sec:nucleo}, we examine the nucleosynthesis predictions used to investigate the nature of Tc-rich M stars. 
The kinematics of Tc-poor, Tc-rich M stars and Tc-rich S stars are compared in Sect.~\ref{sec:kinematics}.
Finally, our results are discussed in Sect.~\ref{sec:discussion}, and the main conclusions of this study are summarized in Sect.~\ref{sec:conclusion}.

\section{Observational sample}\label{sec:sample}
We obtained high-resolution optical spectroscopic data from the HERMES spectrograph \citep{Raskin2011} for a sample of M stars. 
The radial velocity of each spectrum was determined by measuring the barycentre of its cross-correlation function. 
The latter was computed using a line mask designed from a high signal-to-noise HERMES spectrum of an M4 star.
The spectral resolution is R$ = 86\,000$ and the typical signal-to-noise ratio (SNR) of our sample spectra is $30$ in the $V-$ band, providing sufficient quality to detect the technetium lines if present.
%to ensure the detection of technetium. 
However, for five stars, namely U~Per, T~Cep, R~Aqr, RX~Lac and Y~And, the SNR was sufficient for the Tc detection but not for the computation of spectral indices discussed in Sects.~\ref{sec:Tclinefitting} and \ref{sec:TiOZrO}.
The sample stars and their basic properties are presented in Table~\ref{tab:basicdata}.

The selection of this sample was guided by the following specific criteria. 
First, 
we used the list of M stars from \cite{Little1987} as the starting point. From this dataset, only stars meeting the following conditions were included: a declination $\delta \geq -30^\circ$, an apparent magnitude $V \leq 11$ (with the average $V$-magnitude adopted for variable stars), and the availability of a \referee{Gaia data release 2} parallax.
Next, we calculated the bolometric magnitudes 
using Gaia parallaxes and \referee{\textit{Johnson} $V-K$ \citep{Johnson2002}} color indices for all candidate stars.
%using Gaia parallaxes. 
These calculations assumed a standard interstellar reddening value of 0.1 mag and applied a bolometric correction in the $K-$band of 3.0 mag. Finally, based on the $M_{\mathrm {bol}}$ and $V-K$ range of Tc-rich M stars from \cite{Uttenthaler2011}, we established a specific selection region, namely $-6 \leq M_{\mathrm {bol}} \leq -3$ and $5 \leq V-K \leq 12$. This criterion allowed us to identify M stars with a high probability of being Tc-rich.
Indeed, as will be seen in the next section, 46\% of the M star sample turned out to be Tc-rich. 

\section{Confirmation of the Tc presence}\label{sec:tc}
The three Tc I resonance lines located at 4238.19, 4262.27 and
4297.06 \AA~were used to detect the presence/absence of Tc in our sample stars. In Figs.~\ref{fig:Tcrichset1} and \ref{fig:Tcrichset2} (resp., Figs.~\ref{fig:Tcpoorset1} and \ref{fig:Tcpoorset2}), we present the Tc lines of the Tc-rich (resp., Tc-poor) M stars.

We identified \tcrichM~stars as Tc-rich based on the presence of absorption features in either two or all three Tc lines. Among these, R~Vir and Y~And were previously classified as Tc-poor by \cite{Little1987} 
but Fig.~\ref{fig:Tcrichset2} unambiguously proves that they are Tc-rich.
V~Mon, ST~Her, U~Per, T~Cas, and RZ~Her were labeled as "probable" Tc-rich candidates by \cite{Little1987}. Of these, ST~Her was later confirmed as Tc-rich by \cite{Lebzelter1999}, and we confirm the Tc-rich nature of the remaining stars in this work. The other Tc-rich candidates in our sample were already identified as such in previous studies, including \cite{Little1979, Little1986, Smith1988, Jorissen1993, Lebzelter1999, Uttenthaler2011}, and our findings are consistent with those classifications.

Based on the spectral features shown in Figs.~\ref{fig:Tcpoorset1} and \ref{fig:Tcpoorset2}, we classify \tcpoorM~stars in our sample as Tc-poor.
Previously, \cite{Little1987} labeled TV~Psc, RV~Boo, V1351~Cyg, NZ~Gem, and UV~Her as "doubtful", noting that any lines at the Tc I positions were either too weak or heavily blended, suggesting the absence of Tc.
In this study, we confirm these stars as Tc-poor, except for RT~Cyg, where the low SNR of our spectra prevents a definitive classification.
Similarly, \cite{Little1987} classified X~Oph, R~CVn, and V~Cas as "possible", meaning the Tc lines in their spectra were weak or blended, with a preference toward the presence of Tc. However, our analysis shows no clear absorption features at the Tc lines for X~Oph and R~CVn. While the SNR of the V~Cas spectrum is low (Fig.~\ref{fig:Tcpoorset2}), the Tc lines at 4262.27~\AA~and 4297.06~\AA~suggest a Tc-poor classification for this star. 
V~Boo was classified as Tc-rich by \citet{Lebzelter1999}, although they expressed uncertainty about this classification and noted the need for high-resolution data to confirm it. Based on our analysis, we identify the star as Tc-poor.
The remaining Tc-poor stars in our sample were previously identified as such by \cite{Little1987, Smith1988, Lebzelter1999, Lebzelter2003}. 
Because most Tc-rich M stars appear to be cool large-amplitude variables (Table~\ref{tab:basicdata}) with spectral type (and hence temperature) varying over the pulsation cycle, their abundance analysis turned out to be extremely difficult \citep{shetye2022}. Tc (and Zr) abundances could therefore not be derived for any of our sample stars, with the exception of the low-variability MS star AA~Cam (Table~\ref{tab:basicdata}), analysed by \citet{abc2021}.

\section{Technetium line fitting}\label{sec:Tclinefitting}
\begin{figure}
    \centering
    \includegraphics[width=1.05\linewidth]{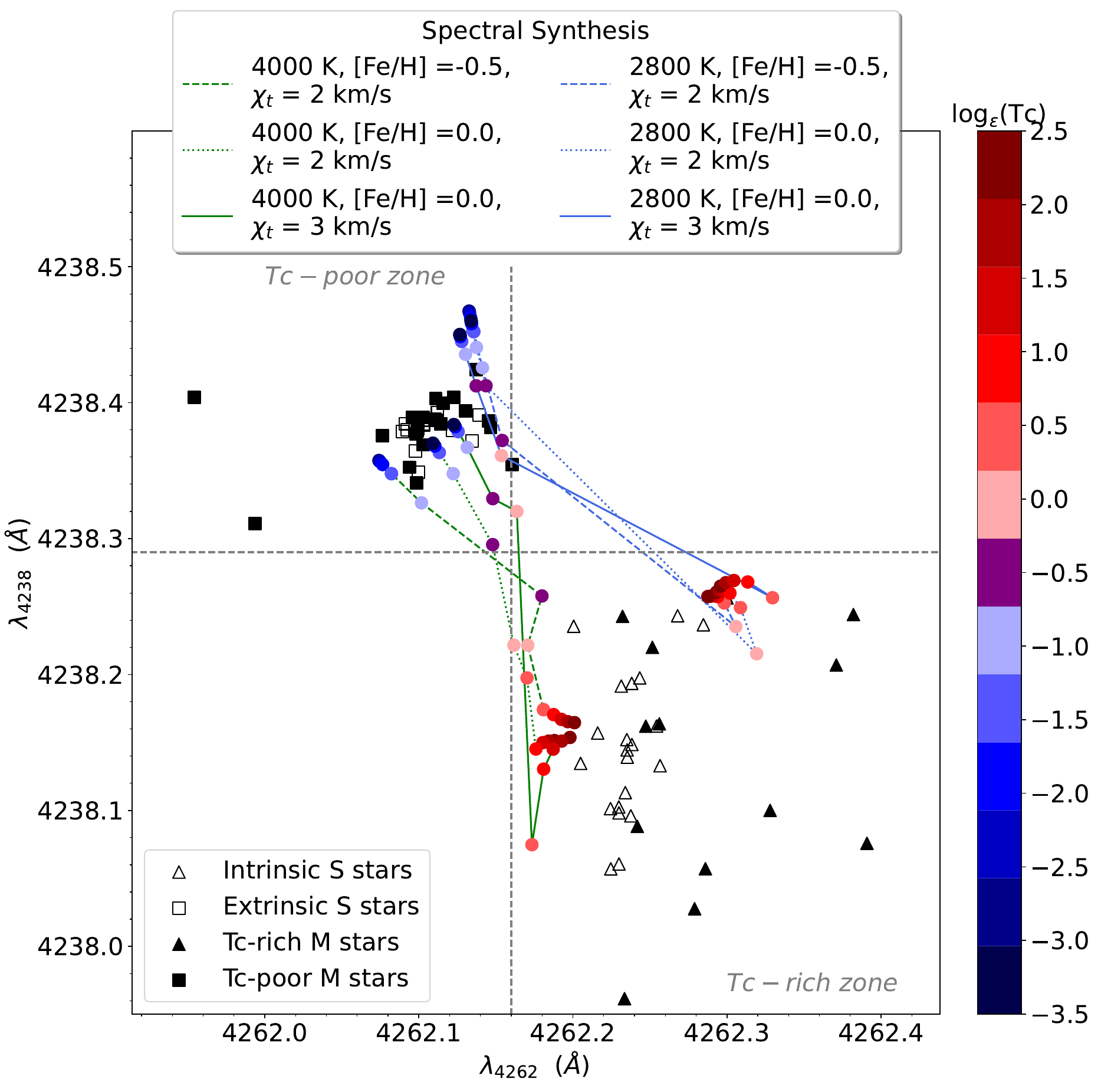}
    \caption{Central wavelengths of the Tc~I blends at $\lambda {4238}$ and $\lambda {4262}$ Tc~I.
    The grey dashed lines delineate the boundaries between the \referee{Tc-poor zone} (upper left rectangle) and the Tc-rich region \cite[lower-right rectangle, ][]{Vaneck1999}.
    Observed data points are shown as squares and triangles, corresponding to measurements of ($\lambda_{4238}$, $\lambda_{4262}$) in spectra of Tc-rich and Tc-poor M- and S-type stars, as labeled.
    Synthetic spectra were generated for two effective temperatures (\teff\ = 2800~K and 4000~K), metallicities ([Fe/H] = 0.0 and -0.5), two microturbulence velocities ($\chi_t = 2$ and $3$~km/s), and a range of technetium abundances (as indicated by the color scale on the side).
    From these synthetic spectra, the central wavelengths of the $\lambda4238$ and $\lambda4262$ Tc~I blends were measured. The results are represented as colored circles, connected by lines that group models with the same atmospheric parameters, as indicated by the labels.}
    \label{fig:Tclinebarycenter}
\end{figure}
Spectral synthesis rarely produces a satisfactory fit at the location of the blue Tc lines. Therefore the presence or absence of the Tc lines determined in Sect.~\ref{sec:tc} was confirmed as described by \cite{Vaneck1999}. 
The shape and position of the blends for the two Tc lines at 4238~\AA\ and 4262~\AA\ exhibit distinct characteristics in Tc-rich and Tc-poor stars. The Nb~I–Gd~II (–Tc~I) blend, referred to here as the $\lambda_{4262}$ feature, is significantly influenced by the presence or absence of the Tc line. Similarly, the CH–La~II blend associated with the Tc line at 4238~\AA\ is designated as the $\lambda_{4238}$ feature. We fitted a Gaussian to these two Tc features for both Tc-rich and Tc-poor stars and derived the wavelength corresponding to the minimum of the fitted Gaussian. According to \cite{Vaneck1999}, the minimum of the $\lambda_{4262}$ feature in Tc-rich stars is redshifted by approximately 0.14~\AA\ compared to a Gaussian fitted to the $\lambda_{4262}$ feature in Tc-poor stars. In contrast, the minimum of the $\lambda_{4238}$ feature shows a blueshift of approximately 0.2~\AA~in Tc-rich stars. 

In Fig.~\ref{fig:Tclinebarycenter}, we present the wavelengths of the $\lambda_{4238}$ feature plotted against those of the $\lambda_{4262}$ feature. The separation between Tc-rich and Tc-poor stars in this plane is strikingly clear. Tc-rich M stars occupy the same region as Tc-rich S stars, while Tc-poor M and S stars are similarly grouped together.  
The two Tc blends (at $\lambda_{4238}$ and $\lambda_{4262}$) always produce consistent results concerning the presence/absence of the Tc lines.
\referee{Close inspection of the Tc-rich M-star properties suggest that the increased scatter in the lower right part of Fig.~\ref{fig:Tclinebarycenter} may be linked to the cooler temperatures and stronger variability (atmospheric dynamics) of those particular Tc-rich stars, which are generally more evolved than their Tc-poor counterparts.}

For comparison, we applied the same Tc-line fitting 
to synthetic spectra. These spectra were generated using the radiative transfer code Turbospectrum \citep{Alvarez1998, Plez2012} combined with MARCS model atmospheres for S-type stars from \cite{VanEck2017}. Synthetic spectra were computed for varying Tc abundances, spanning \logTc~from $-3.5$ to $2.5$, and for two extreme temperatures from the MARCS grid of S stars, namely 2800~K and 4000~K. We also considered two metallicities, [Fe/H]~=~0.0 and $-0.5$, and investigated the impact of microturbulence by computing synthetic spectra for $\chi_t= 2$ and $3$~km/s.  
The MARCS models of S-type stars from \cite{VanEck2017} provide a selection of C/O and [s/Fe] ratios covering the typical s-process and carbon abundances observed in S-type stars. 
For our analysis, we used synthetic models with a fixed C/O ratio of 0.5 and [s/Fe] = 0 dex to closely match the composition of M-type stars.

From the synthetic spectra, the wavelength of the minimum for both features exhibits a clear dependence on effective temperature. 
Metallicity, in turn, impacts the minimum Tc abundance for the star to be classified as Tc-rich. 
Notably, a step of 0.5~dex in \logTc~is sufficient to shift a star from the Tc-poor zone to the Tc-rich zone. 
The smallest Tc abundance for Tc-rich stars is $\log \epsilon(\mathrm{Tc}) = -0.5$ and occurs for spectral synthesis computed with \teff = 4000K and [Fe/H] = $-0.5$.
These detection limits may be 
only appropriate for M-type stars. 
Tc-rich S-type stars exhibit Tc features even at relatively low Tc abundances (as low as \logTc~$\sim$~ $-1.3$, see S-type stars in Fig.~\ref{fig:nucleopredictions}). 
Consequently, the Tc barycenter may also be sensitive to 
increase in the C/O ratio or [s/Fe], through the s-process element blends affecting the Tc~I lines.

\section{TiO-ZrO band indices}\label{sec:TiOZrO}
\begin{figure*}
    \centering
    \includegraphics[width=1.0\linewidth]{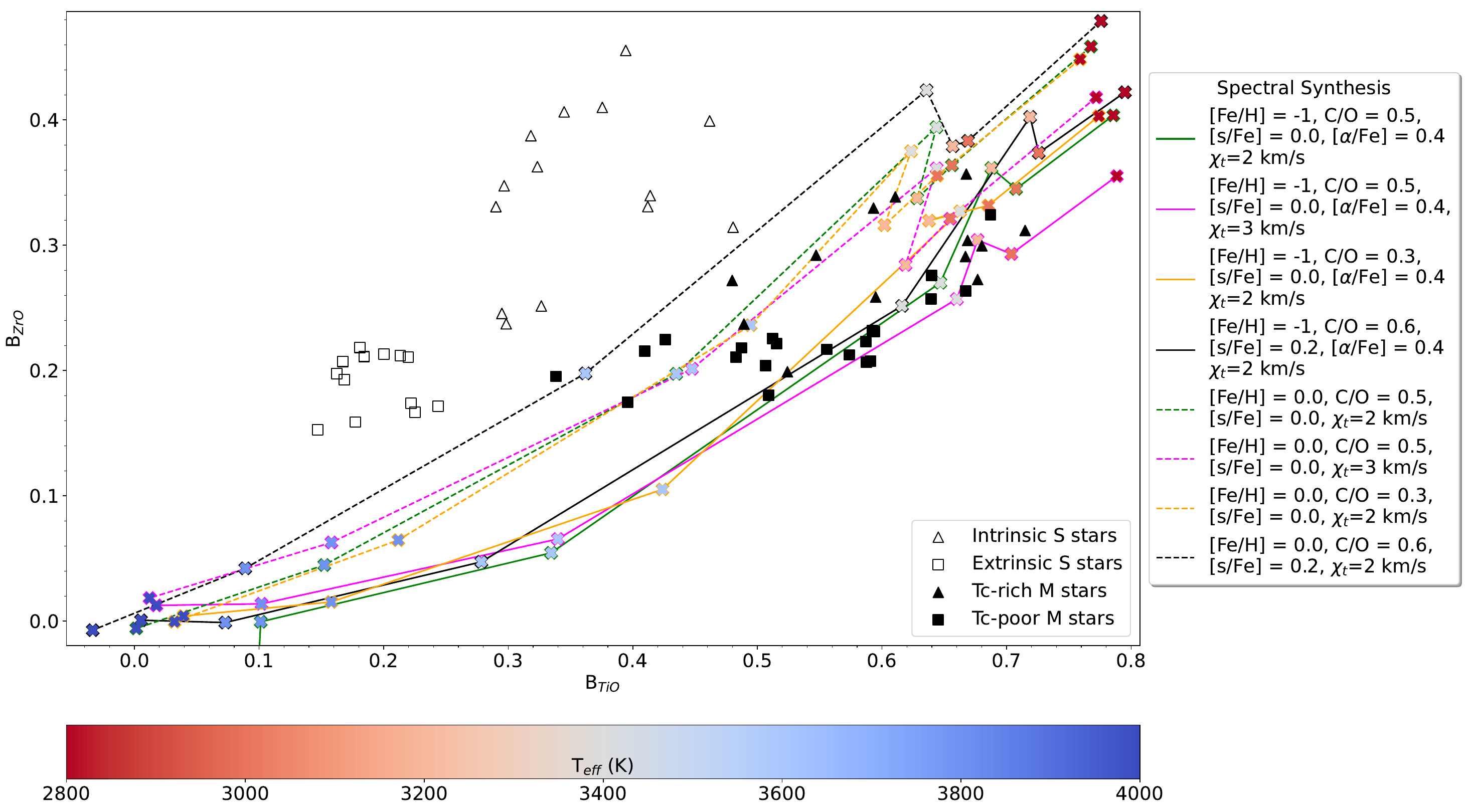} %trim={4cm 2cm 0 1cm}
    \caption{The band strength indices $B_{\TiO}$  and $B_{\ZrO}$ are shown for Tc-rich and Tc-poor M-type stars from this study, as well as for Tc-rich (intrinsic) and Tc-poor (extrinsic) S-type stars from \citet{abc2021}. 
    They are represented by filled or open squares and triangles, as indicated in the legend.
    Corresponding spectral indices were also computed from synthetic spectra based on model atmospheres with varying stellar parameters, including effective temperature (color-coded according to the colour bar), metallicity, C/O ratio, s-process enrichment [s/Fe], alpha-elements enrichment [$\alpha$/Fe] and microturbulence $\chi_t$.
    Band strengths derived from synthetic spectra are shown as crosses, with models of [Fe/H]$= 0$ connected by dashed lines and those with [Fe/H]$=-1$ connected by solid lines.}
    \label{fig:ZrOTiO}
\end{figure*}

We calculated band-strength indices for the TiO and ZrO bands, denoted as $B_{\TiO}$ and $B_{\ZrO}$, respectively, for both observed and synthetic spectra. Details on the computation and on the continuum and band windows used for these indices are provided in \cite{VanEck2017}.
\referee{In summary, we used four ZrO bands within the wavelength range of 5740–6530 \AA~and seven TiO bands between 5440 and 7150 \AA.
B$_{\TiO}$ and B$_{\ZrO}$ were calculated for each molecular band using Equation 1 of \cite{VanEck2017}. 
The final index for each molecule was derived by averaging the values obtained from the individual bands.}

These spectral indices indicate that
M stars generally exhibit significantly stronger TiO absorption bands compared to both Tc-poor and Tc-rich S stars
while their ZrO bands are comparable in depth to those of Tc-rich S stars.   
A comparison of Fig.~\ref{fig:ZrOTiO} with Figs.~6 and 8 of \cite{VanEck2017} demonstrates that M-, S-, and C-type stars occupy distinct locations in the $B_{\TiO}$–$B_{\ZrO}$ plane. 
In addition to the temperature effects,
this separation is driven by the strong dependence of ZrO and TiO band strengths on the C/O ratio: varying the C/O ratio produces a clear displacement of the synthetic band indices within this diagram.

Within the S star sample, 
intrinsic S stars exhibit significantly stronger ZrO and TiO bands, likely due to their cooler temperatures compared to extrinsic S stars. Similarly, among the M stars in the current study, a separation between Tc-rich and Tc-poor M stars can be observed, albeit less clearly than for S stars. Interestingly, four Tc-poor M stars (V~Cas, UV~Her, R~CVn, R~Leo) show ZrO band strengths comparable to those of Tc-rich M stars. These stars are highly variable, suggesting that they are cooler than other Tc-poor M stars, which could account for their enhanced ZrO and TiO band strengths.

To further investigate the dependence of the $B_{\TiO}$ and $B_{\ZrO}$ indices on stellar parameters, we compared them with indices computed from synthetic spectra. In Fig.~\ref{fig:ZrOTiO}, we present the synthetic band indices as a function of varying effective temperature, metallicity, C/O ratio, [s/Fe]
\footnote{\referee{Logarithmic abundance ratio of s-process elements to iron, relative to the solar ratio. It quantifies the enrichment of s-process elements (typically in the range Z=31 to Z=83) compared to iron.}}
and microturbulence of the models used to compute the spectra. 
As expected, models with \teff$\approx 4000$~K produce $B_{\TiO} \approx B_{\ZrO} \approx 0$, as expected for K giants with very weak molecular bands. 
The models that best reproduce the  Tc-rich M star TiO and ZrO band strengths are [Fe/H]~=~$-1$ models. Agreement is also found for some stars for the [Fe/H]~=~0 models with C/O $\leq 0.5$, which is actually the range expected for M-type stars.

\section{Stellar parameters and the HR diagram}\label{sec:stellarparam}
\subsection{Metallicity [Fe/H] derivation of OP~Her and RX~Lac}\label{sec:meta}
To estimate the metallicity of the sample stars, we conducted an abundance analysis on the two least variable Tc-rich M stars in our sample: OP Her ($\Delta V$ = 0.90~mag), and RX Lac ($\Delta V$ = 1.54~mag).
The initial effective temperature estimates were obtained from the ZrO-TiO analysis (see Fig.~\ref{fig:ZrOTiO}). Using these estimates as a starting point, we compared the observed spectra of these stars with synthetic spectra generated at the same temperature. The synthetic spectra were produced using MARCS model atmospheres for S-type stars \citep{VanEck2017} and Turbospectrum \citep{Plez2012}. We iteratively adjusted the C/O ratio and titanium abundance to optimize the agreement between the observed and synthetic spectra. 
\referee{A satisfactory match was defined as a good reproduction of key molecular bands, such as ZrO, TiO, and CN. Once this was achieved,}
we analyzed 6 to 12 Fe lines to determine the metallicity of the two stars. The stellar parameters used for this analysis and the derived metallicities are presented in Table~\ref{tab:meta}.
The typical uncertainties in our derived parameters are $\pm$100 K for the effective temperature, $\pm$1 dex for surface gravity (log~g), and $\pm$0.25 for the C/O ratio (for more details on typical uncertainties in evolved stars, see \citealt{shetye2018, abc2021}). 
The uncertainties in luminosity are listed in Table~\ref{tab:meta} and were calculated based on the errors in the \textit{Gaia} DR3 parallaxes.
For comparison, we also included the stellar parameters and metallicity of AA~Cam (an MS star with $\Delta V$ = 0.24~mag) from \cite{abc2021}.

Although based on this limited sample, there is evidence that the metallicities of Tc-rich M stars are subsolar
with [Fe/H] values of $-0.3$, $-0.6$, and $-0.9$ (see Table~\ref{tab:meta}).
The average metallicity derived for the three Tc-rich M (or MS) stars is slightly lower than that of the Tc-rich S star sample, which exhibits a [Fe/H] distribution peaking at $-0.3$ \citep{abc2021}. It is important to emphasize that the abundance analysis of the Tc-rich M stars in this sample is particularly challenging, as discussed in detail by \cite{shetye2022}. Consequently, a comprehensive abundance analysis of the full sample is deferred to a forthcoming paper.

\subsection{Location of AA~Cam, OP~Her and RX~Lac in the HR diagram}\label{sec:HRD}
\begin{figure*}
    \centering
    \includegraphics[width=0.465\linewidth]{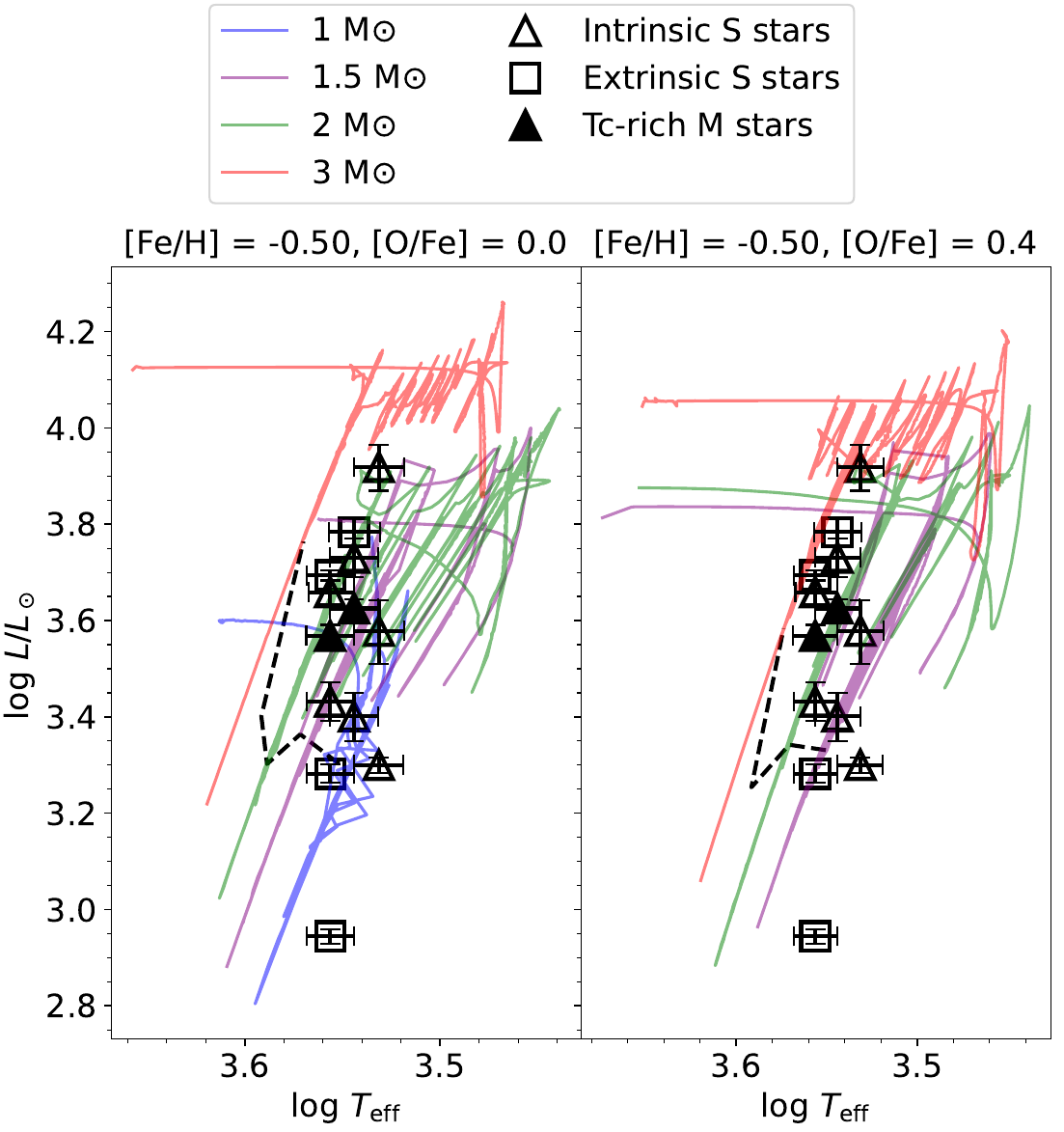}
    \includegraphics[width=0.467\linewidth]{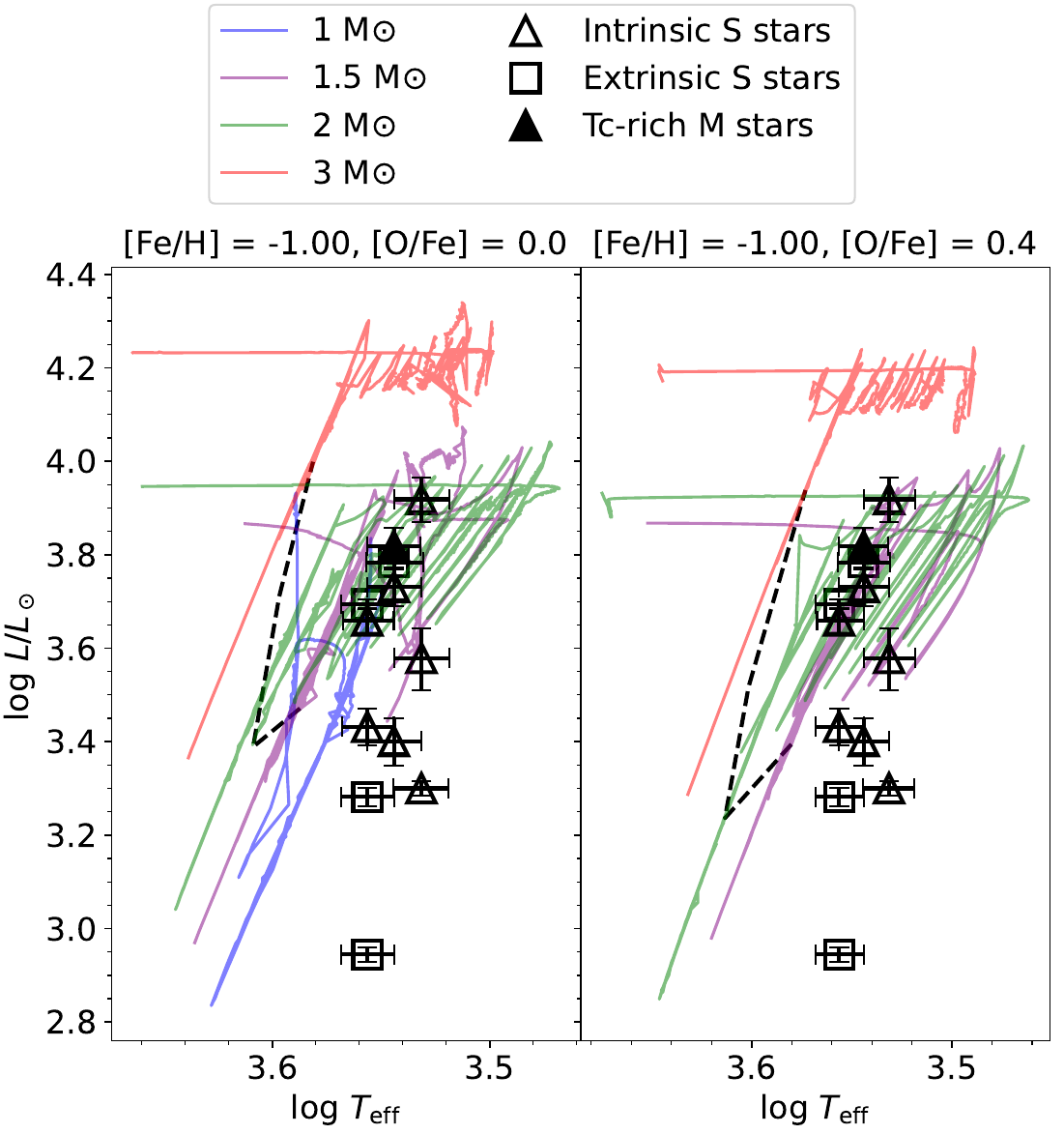}
    \caption{Locations of the Tc-rich M stars AA~Cam, OP~Her and RX~Lac (filled triangles) in the HR diagram.
    \referee{The HR diagrams also include STAREVOL evolutionary tracks corresponding to the closest available metallicities ([Fe/H] = –0.5 or –1.0) with [O/Fe] = 0 and 0.4, hence AA Cam and OP Her appear in the left subpanels, while RX Lac is shown in the right subpanels. }
    The black dashed line represents the predicted onset of the TDU, that is, the lowest stellar luminosity following the first occurrence of a TDU episode. For comparison, Tc-rich S stars (empty triangles) from \cite{abc2021} and Tc-poor S stars (empty squares) from \cite{shetye2018}~with metallicities around [Fe/H]~$=-0.5$ are included. }
    \label{fig:HRD}
\end{figure*}

A tentative mass estimate for some Tc-rich M stars can be obtained by placing them in the HR diagram (Fig.~\ref{fig:HRD}). However,
 their intrinsic
variability in luminosity and effective temperature introduces
significant uncertainties. This variability complicates the direct
comparison of their locations in the HR diagram with the STAREVOL evolutionary tracks \citep{Siess2000}.
For the least variable Tc-rich M stars (for which we derived metallicities in the previous section - see Table~\ref{tab:meta}), we 
calculated their luminosities using the method outlined in \cite{shetye2018,abc2021}. This approach utilizes parallaxes from $Gaia$ Data Release~3 \citep{GaiaDR32023} and extinction values from the \cite{Gontcharov2017} reddening map. 
These three Tc-rich M stars (namely AA~Cam, OP~Her, and RX~Lac) are all located beyond the predicted onset of the TDU (indicated by the black dashed line in Fig.~\ref{fig:HRD}) and overlap with the region occupied by Tc-rich S stars. 
Thus AGB stars do not strictly adhere to the canonical M-S-C evolutionary sequence in the HR diagram. 
A similar finding was reported by \citet{Uttenthaler2024}, who observed comparable luminosity functions for Tc-rich M, MS, and S stars, 
and this finding adds further complexity to the
understanding of how these stars evolve.

Because stars of metallicity [Fe/H] $=-1$ generally exhibit enhanced [O/Fe] by 0.4~dex, we also compared the positions of our Tc-rich M stars with oxygen-rich evolutionary tracks (see Sect.~\ref{sec:nucleo} for details about these models).
However, these O-rich tracks do not favour Tc-rich M masses lower than those predicted by the standard tracks, even though masses as low as 1~\msun\ seem to be required by their kinematical properties, as it will be concluded from Sect.~\ref{subsec:kine} below.

From Fig.~\ref{fig:HRD} we can conclude that
the positions in the HR diagram of the three Tc-rich M stars studied are consistent with low-metallicity $1.5 - 2$~M$_\odot$ evolutionary tracks (and perhaps even down to masses $\sim1.0$~M$_\odot$, as formerly concluded by \citealt{shetye2019} for Tc-rich S stars with the lowest luminosities). Moreover, the 3~M$_\odot$ tracks can be ruled out.

\begin{table}[]
\caption{Stellar parameters and metallicity of some sample stars.}
    \label{tab:meta}
    \centering
    \addtolength{\tabcolsep}{-0.2em}
    \begin{tabular}{c|cccccc}
    \hline
    \hline
    Star & T$_{\rm {eff}}$ & $L$ & $\log~g$ & [Fe/H] & $\sigma_{\rm{[Fe/H]}}$ & C/O \\
        & (K) & ($L_\odot$) &        & (dex)  & (dex)             &      \\
    \hline
       AA Cam  & 3600 & $3700${\raisebox{0.5ex}{\tiny$^{+200}_{-100}$}} & 1.0 
       & $-0.3$ (13) & 0.13  & 0.5\\
       OP Her  & 3500 & $4200${\raisebox{0.5ex}{\tiny$^{+200}_{-100}$}} & 0.0 
       & $-0.6$ (8) & 0.32  & 0.3\\
       RX Lac  & 3500 & $6600${\raisebox{0.5ex}{\tiny$^{+700}_{-600}$}}&  0.0 
       & $-0.9$ (7) & 0.16 & 0.4 \\
    \hline
    \hline
    \end{tabular}
 \tablefoot{The stellar parameters and abundances of AA~Cam are retrieved from \cite{abc2021}. The metallicities [Fe/H] have been obtained from the Fe abundance analyses. The numbers in brackets in the [Fe/H] column indicate the number
of lines used to derive [Fe/H] and the next column indicates the standard deviation (derived from the line-to-line scatter) on [Fe/H].}
\end{table}

\section{Nucleosynthesis predictions}\label{sec:nucleo}
\begin{figure}
    \centering
    \includegraphics[width=1.0\linewidth]{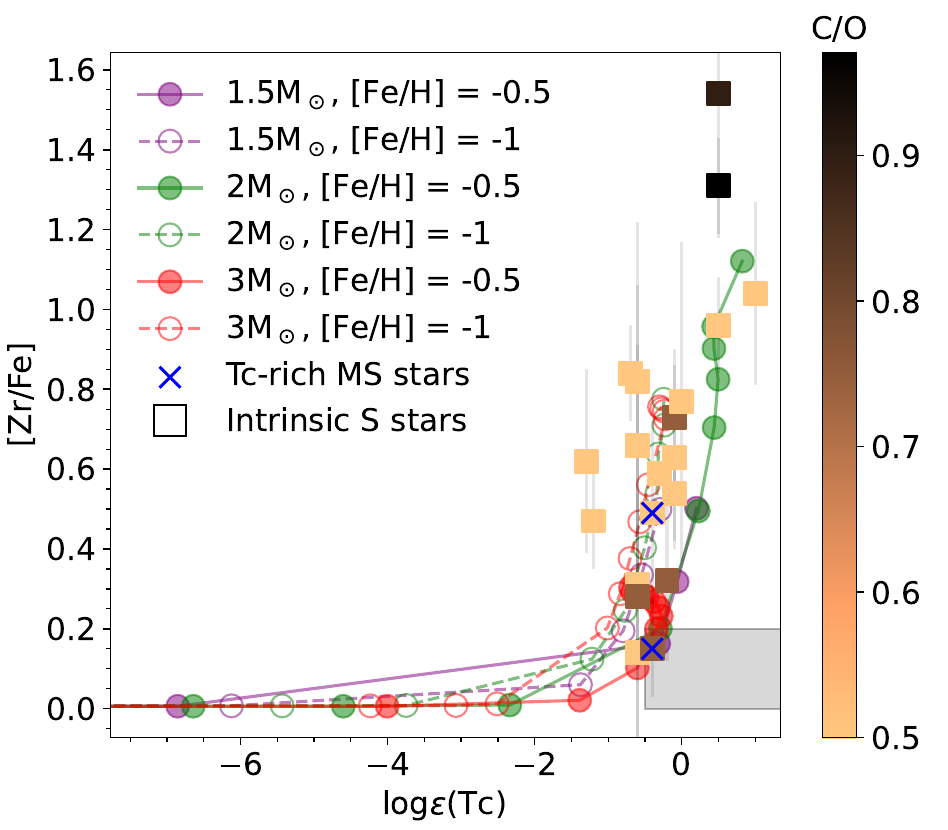}
    \caption{ Nucleosynthesis predictions for Tc and Zr abundances during successive TDU events are shown as circles along the model tracks. Measured abundances in Tc-rich S stars (filled squares), from \cite{abc2021}, are color-coded by C/O ratio. 
Blue crosses highlight the two Tc-rich MS stars (AA~Cam and $\omicron^1$~Ori) in their sample.
The grey box highlights the expected locus of Tc-rich M stars, where the Tc range reflects the detection threshold for Tc-rich classification (see Fig.~\ref{fig:Tclinebarycenter}), and the upper limit of the Zr range corresponds to the black dashed line in Fig.~\ref{fig:ZrOTiO} with [s/Fe]~=~0.2, which separates M-type from S-type stars. 
    \label{fig:nucleopredictions}
    }
\end{figure}

Stellar models with initial mass $M=$~\{1.5, 2 , 3\}~\msun{} and metallicities [Fe/H] = \{-0.5, -1\} were computed with the STAREVOL code \citep{Siess2000}. We recall its main physical ingredients~: the nuclear network includes 414 nuclei and 749 nuclear reactions ($\beta-$ decay, $n-$, $p-$, $\alpha-$  and electron captures) as described in \cite{Goriely2018}. Prior to the thermally pulsing AGB phase, we use the \cite{Schroder2007}  mass loss rate prescription and then take the maximum between \cite{Schroder2007} and \cite{Vassiliadis1993}. We also consider core overshooting following the mass dependence relation of \cite{Claret2018} and during the AGB phase extra-mixing below the convective envelope is modeled as in \cite{Goriely2018} with the following mixing parameters: $p=5$, $D_\mathrm{min} = 10^9$ and $f_\mathrm{over} = 0.1$. We use the solar composition of \cite{Asplund2009} and mixing length parameter $\alpha_\mathrm{MLT} = 1.75$. 
For the O-rich models, we use the same solar scale composition as before except that we impose [O/Fe] = 0.4. The excess mass fraction due to this enrichment is then deduced from the H mass fraction. 

Figure~\ref{fig:nucleopredictions} shows the predicted nucleosynthesis trends for the evolution of Tc and Zr abundances over successive thermal pulses 
and TDU events during the AGB phase.
During the initial 
TDU events, the Tc abundance, that we trace via \logTc
\footnote{The logarithmic abundance of technetium, denoted as \logTc, is defined as: \[ \log \epsilon(\mathrm{Tc}) = \log_{10} \left( \frac{N_{\mathrm{Tc}}}{N_{\mathrm{H}}} \right) + 12\] where $N_{\mathrm{Tc}}$ and $N_{\mathrm{H}}$ are the number densities of technetium and hydrogen atoms, respectively. The constant 12 is a normalization factor that sets the abundance of hydrogen to $\log \epsilon(\mathrm{H}) = 12.00$ by definition.}, 
increases with no observational effect on the Zr abundance. 

A slight [Zr/Fe] enhancement emerges once $\log \epsilon$(Tc) reaches between $–1.5$ and $-0.5$ depending on the model.
This delayed increase in Zr results from an existing non-zero (actually solar-scaled) Zr abundance:
since $\rm{[Zr/Fe]} = \log \epsilon(Zr) - 2.58 -  \rm{[Fe/H]}$
\citep[where $\log\epsilon(Zr)_\odot = 2.58$, ][]{Asplund2009},
as long as the Tc increase $\Delta N$(Tc) is of the order of $10^{-6}$ to $10^{-2}$, the expected similar increase in 
$\Delta N$(Zr) is not noticeable on [Zr/Fe].
Consequently, the spectral change at the location of the Tc line is very noticeable, whereas the increase in the ZrO band depth is more progressive.
%and discrete.
This gradual increase in [Zr/Fe] contrasts with the more abrupt Tc increase from zero, providing a natural explanation for Tc-rich M stars.

Since no successful abundance analyses  could be performed for Tc-rich M stars (see Sect.~\ref{sec:tc}), the few stars plotted in Fig.~\ref{fig:nucleopredictions} are Tc-rich S stars from \citet{abc2021}, which include two Tc-rich MS\footnote{MS stars are stars sharing the distinctive spectral properties of M stars (TiO bands) and S stars (ZrO bands).}  stars (AA~Cam and $\omicron^1$~Ori; blue crosses), which are much less variable than Tc-rich M stars, hence much easier to analyse.

The classification thresholds between Tc-poor and Tc-rich stars, and between M- and S-types, define a narrow region in Fig.~\ref{fig:nucleopredictions} where Tc-rich M stars are expected to lie.
Based on the tests performed on synthetic spectra as summarized in Fig.~\ref{fig:Tclinebarycenter}, 
stars are typically identified as Tc-rich only when
 $\log\epsilon(\mathrm{Tc}) > -0.5$. 
Conversely, the transition from M- to S-type occurs near
$\rm{[Zr/Fe]} = 0.2$ 
\citep[see Fig.~6 of][as well as the black dashed line corresponding to the s-process index of 0.2 dex in Fig.~\protect\ref{fig:ZrOTiO}]{abc2021}.
This small region is delineated by the grey rectangle in Fig.~\ref{fig:nucleopredictions}.
Actually, most of the nucleosynthesis models considered fall outside the Tc-rich M box. 
Only [Fe/H]~$= -0.5$ models
are marginally  compatible with this Tc-rich M box. 
Whereas one of the Tc-rich stars, $\omicron^1$ Ori, lies in the grey region, the other, AA~Cam, does not as its Zr abundance is comparable to that of intrinsic S stars. 
However, since AA~Cam was classified as M5S after its Zr enhancement was identified, it cannot be considered a genuine, prototypical Tc-rich M star.

\section{Spatial and kinematical properties}\label{sec:kinematics}

\begin{table*}[]
\setlength{\tabcolsep}{3.0pt}
    \centering
    \caption{Spatial and kinematical properties of the samples, roughly ordered from younger to older (according to $\sigma_u$), from left to right (excluding the data from the Besançon Galactic Model). %\LS{how are these numbers determined?}
    }
    \label{tab:velo}
    \begin{tabular}{cc|rrrrrrrrr|rrrrrr}
    \hline
    \hline\\
  & & \multicolumn{1}{c}{Intrinsic} & \multicolumn{1}{c}{N-type} & \multicolumn{1}{c}{Extrinsic} && \multicolumn{1}{c}{giant} &&  \multicolumn{1}{c}{Tc-poor} & \multicolumn{1}{c}{Tc-rich} && 5.4 - 2.5  & 1.9 - 1.6 & 1.6 - 1.3 & 1.3 - 1.15 & 1.15 - 1 & M$_{\odot}$\\
   &       & \multicolumn{1}{c}{S}   & \multicolumn{1}{c}{C}   &  \multicolumn{1}{c}{S$^a$}    && 
   \multicolumn{1}{c}{M} &&
   \multicolumn{1}{c}{M}       & \multicolumn{1}{c}{M}  &&  \multicolumn{1}{c}{0.15 - 1} & 
   \multicolumn{1}{c}{2 - 3} & \multicolumn{1}{c}{3 - 5} & \multicolumn{1}{c}{5 - 7} & \multicolumn{1}{c}{7 - 10} & Gyr\\   \\
          \cline{3-5}\cline{7-7}\cline{9-10}\cline{12-17}
    &         &\multicolumn{3}{c}{Abia}&& \multicolumn{1}{c}{Famaey}& &  \multicolumn{2}{c}{This work}&&\multicolumn{5}{c}{Besançon Galactic Model}\\
         \hline
&
$\sigma_u$ (km/s) &27.4 & 27.7 & 33.1 && 31.8~$\pm$1.6
&&38.8~$\pm$~0.5  & 46.3~$\pm$~0.7 &&  \multicolumn{1}{c}{12.7} & 
 \multicolumn{1}{c}{24.4} &  \multicolumn{1}{c}{30.4}&
  \multicolumn{1}{c}{36.9}&
   \multicolumn{1}{c}{42.7}\\
 & $\sigma_v$ (km/s)&18.2 & 19.0 & 24.8 && 17.6~$\pm$0.8
 &&19.7~$\pm$~0.2  & 29.4~$\pm$~1.3 &&  \multicolumn{1}{c}{7.2}  & 
  \multicolumn{1}{c}{13.9} &  \multicolumn{1}{c}{17.2}&
   \multicolumn{1}{c}{21.0}&
    \multicolumn{1}{c}{24.3}\\
 & $\sigma_w$ (km/s)&15.2 & 12.6 & 16.3 && 
 16.3~$\pm$0.8
 &&18.6~$\pm$~0.4  & 12.3~$\pm$~0.3 &&  \multicolumn{1}{c}{5.9}  & 
  \multicolumn{1}{c}{11.3} & 
   \multicolumn{1}{c}{14.0}& 
    \multicolumn{1}{c}{17.0} & 
     \multicolumn{1}{c}{19.8} &
\medskip\\
 & $H_Z$ (pc) &  150   & 200 & 260 &&
 \multicolumn{1}{c}{196}
 &&\multicolumn{1}{c}{190} & \multicolumn{1}{c}{190} \\
& $N$ & 50 &144&85$^b$&& 
\multicolumn{1}{c}{3417}
&&
\multicolumn{1}{c}{21}& \multicolumn{1}{c}{18}\\
       \hline
       \hline
    \end{tabular}
    \tablefoot{$\sigma_u, \sigma_v, \sigma_w$ are the standard deviations of velocities in the heliocentric Galactic 
coordinate system, $H_Z$ is the Galactic scale height, and $N$ is the number of stars in the sample. Predictions from the Besançon Galactic Model are shown for comparison (as a function of age and main-sequence mass) for the thin disc ($-0.1 \le \mathrm{[Fe/H]} \le 0$), from \cite{Robin2021}. The relation between solar-metallicity main-sequence lifetime $\tau_{\mathrm{MS}}$ and mass $M$ is derived from the simple relation $\tau_{\mathrm{MS}} = 10^9\; (M/M_{\odot})^{-2.5}$ Gyr. See text for more detailed estimates involving low-metallicity AGB stars.\\
$^a$ No correction has been applied to use the binary's center-of-mass velocity, instead of the average velocity.\\
$^b$ Excluding star CD -54$^{\circ}$9245 which behaves like an outlier.}
\end{table*}
\begin{figure}
    \centering
    \includegraphics[width=1.0\linewidth]{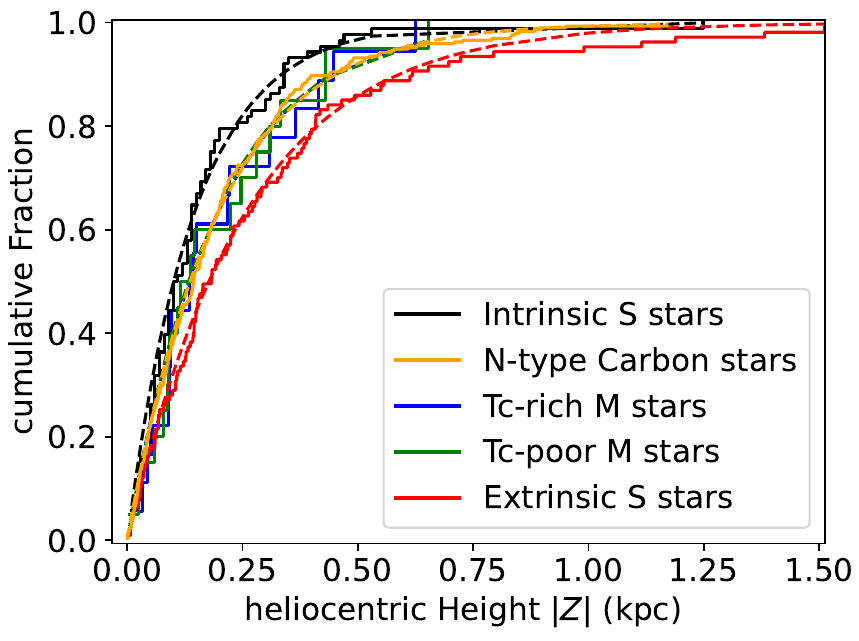}
    \caption{Cumulative distribution of the height above the Galactic plane |Z| (in kpc). The intrinsic and extrinsic S star samples are taken from \cite{Abia2022}, as are N-type carbon stars. The solid lines correspond to the observed distributions and the dashed line to the fit corresponding to an exponentially-decreasing spatial $|Z|$ distribution. The meaning of colors are given in the insert. 
    }
    \label{fig:Z_CDF}
\end{figure}

In this section, we compare the spatial and kinematical properties of Tc-rich and Tc-poor M stars to simulations from the Besançon Galactic Model \citep{Robin2021} as well as to observed properties of other populations of late-type stars, like carbon N stars and Tc-rich or Tc-poor S stars \citep{Abia2022} or the population of kinematically-normal M giants defined by \citet{Famaey2005}, that is, M giants not belonging to any of the streams identified by the latter authors.  
We adopt the Bayesian distances provided by \cite{BJdistances2021}, which are based on a Bayesian analysis of \textit{Gaia}~EDR3 parallaxes. 
R~Leo is excluded from the kinematical and spatial analysis as there is no \textit{Gaia}~DR3 parallax available for it.

Galactic scale heights are discussed in Sect.~\ref{subsec:Z}, then kinematics in Sect.~\ref{subsec:kine}, with a summary presented in Sect.~\ref{subsec:synthesis}.

\subsection{Galactic scale heights}
\label{subsec:Z}

In Fig.~\ref{fig:Z_CDF}, 
we compare the heliocentric height above the Galactic plane ($|Z|$) of these populations. 
There is a clear trend of increasing scale height $H_Z$ [derived from the fit of an exponentially decreasing function $\exp(-|Z|/H_Z)$ to the spatial density] in the sequence of Tc-rich S,  Tc-rich M, Tc-poor M, N-type
C, and Tc-poor S stars, the three middle populations being indistinguishable from each other.
Their respective 
scale heights $H_Z$ are listed in the lower part of Table~\ref{tab:velo}.

Their distribution across the Galactic plane suggests that Tc-rich and Tc-poor M stars belong to the same population (but see Sect.~\ref{subsec:kine} below where we argue that the scale height has a limited sensitivity as compared to kinematics). 
In contrast, both Tc-rich and Tc-poor M stars have a larger scale height than Tc-rich S stars, pointing to a younger (that is, more massive?) population for the latter.
On the contrary, Tc-poor S stars are the most dispersed across the Galactic midplane, with the largest scale height (260~pc) of all the samples considered here. This value is comparable to that ($\sim320$~pc) found by \citet{abc2021}, which means that this population is the oldest among the ones considered here. This conclusion is not surprising, as Tc-poor S stars (also known as extrinsic S stars) host a white dwarf companion. Not all these conclusions will be confirmed however from the kinematical analysis presented in Sect.~\ref{subsec:kine}.

\subsection{Kinematics}
\label{subsec:kine}

The kinematical properties of the samples listed in Table~\ref{tab:velo} refer to the heliocentric velocities $(u,v,w)$ in the Galactic 
coordinate system, that is, 
\begin{eqnarray}
\label{Eq:UVW}
u &=& V_R - U_{\odot} \nonumber\\
v &=& V_p - V_{\odot} - V_{\rm LSR} \\
w &=& V_Z - W_{\odot},\nonumber
\end{eqnarray}
where $V_R, V_p, V_Z$ are the stellar velocities in the Galactocentric reference frame 
\citep[as given by][for the N, extrinsic S and intrinsic S stars]{Abia2022}, $V_{\rm LSR}$ \citep[240~km/s, as adopted by][from \citealt{2014Reid}]{Abia2022} is the velocity of the Local Standard of Rest (LSR) in the direction of Galactic rotation, 
and $U_{\odot}, V_{\odot}, W_{\odot}$ are the components of the solar motion in the LSR \citep[11.1, 12.2, 7.2 km/s respectively, as 
adopted by][from \citealt{Schonrich2012}]{Abia2022}.  The conversion from the $V_R, V_p, V_Z$ values given by \cite{Abia2022} to the $u,v,w$ values listed 
in Table~\ref{tab:velo}  has been performed with Eqs.~\ref{Eq:UVW} and the kinematical constants listed above. In the present paper,
the $u,v,w$ velocities have been computed with radial velocities, proper motions
and parallaxes from Gaia DR3 \citep{GaiaDR32023}. 
In cases where Gaia DR3 radial velocities %from \cite{GaiaDR32023} 
were unavailable, we used values derived from their Hermes spectra  using the cross-correlation method \citep{Raskin2011}, corrected from the barycentric velocity.
The uncertainties on 
$\sigma_u,\sigma_v,\sigma_w$ as listed in Table~\ref{tab:velo} are estimated using a Monte Carlo method, sampling across the error distributions of radial velocity, parallax, and proper motion.

The property $\sigma_u
> \sigma_v > \sigma_w$
should be met by the kinematical quantities listed in Table~\ref{tab:velo} \citep[e.g.,][]{Binney2021}. 
This appears indeed to be the case, which constitutes a sanity check of the data, especially in view of the fact that the binarity of Tc-poor S stars may alter the evaluation of their velocity dispersions\footnote{The estimate of the standard deviations  $\sigma_u, 
\sigma_v,$ and $\sigma_w$ by \citet{Abia2022} does not consider the center-of-mass velocity, as it should, but uses instead a velocity randomly selected on the velocity curve depending on the epoch of observation. Adopting a typical semi-amplitude $K$ of 7~km/s for the S-star velocity curves \citep{Jorissen2019}, one may roughly estimate the uncertainty introduced by computing  $\sigma_u, 
\sigma_v,$ and $\sigma_w$ from a  velocity randomly sampled on the velocity curve instead of from the center-of-mass velocity. This uncertainty is on the order of $\sigma - (\sigma^2 - K^2)^{1/2}$, or 1.5 km/s for $\sigma = 15.2$~km/s or 0.9~km/s for $\sigma = 27.4$~km/s, negligible thus.}.

As already inferred above from their spatial properties, the kinematics of Tc-poor M stars qualifies them as older than Tc-rich S stars, which appears to be the youngest \referee{(that is,} most massive) population among the three discussed so far. However, Tc-rich M stars now appear to be much older than both Tc-poor M stars {\it and} Tc-poor S stars. 
Moreover, from a kinematical viewpoint, Tc-rich and Tc-poor M stars are no more equivalent, in contrast to what was  inferred from their identical scale height. It thus appears that kinematical properties offer a more sensitive probe than the scale height. Tc-poor M stars appear as well slightly older than the kinematical reference sample of M giants from \citet{Famaey2005}. 

A comparison between the predictions of the Besançon Galactic Model \citep{Robin2021}, listed in the right-half of Table~\ref{tab:velo}, allows us to be more quantitative, and associate ages to the various populations under consideration.
This comparison reveals that Tc-rich M stars have kinematical properties similar to the oldest stars from the thin disc, aged from 7 to 10~Gyr, corresponding to the main-sequence lifetime of  1 to 1.15 M$_\odot$ solar-metallicity stars. STAREVOL models have been used to evaluate the ages appropriate for evolved low-metallicity AGB stars ($-0.9 \le $[Fe/H]$ \le -0.3$; see Table~\ref{tab:meta}), as are Tc-rich M giants supposed to be.
The kinematics of Tc-rich M stars is just marginally compatible with stars with ages $\sim7$ Gyr (Table~\ref{tab:velo}), that is, low-metallicity ([Fe/H] = $-1$) AGB stars with initial masses of about 1.0~\msun.
As shown by Table~\ref{tab:velo}, in no case however could the Tc-rich M kinematic be reconciled with low-metallicity progenitors of either 2 or 1.5~M$_\odot$ initial masses, which need to be on the order of only 1 to 2 Gyr, respectively, to reach the AGB. 

\subsection{Summary}
\label{subsec:synthesis}
The conclusions that emerge from the spatial and kinematical properties of the various samples studies are not clear-cut, since they are not always consistent with each other. For instance, the Galactic scale heights suggest that Tc-rich and Tc-poor M stars belong to the same population, whereas the kinematics suggests that Tc-rich M stars are older than Tc-poor M stars.
Kinematics suggests as well that Tc-rich M stars are older than Tc-poor S stars, whereas 
Galactic scale heights suggest the opposite.  Kinematics suggests that Tc-rich M stars is the oldest population among the three, whereas from Galactic scale heights, one would instead conclude that the population of Tc-poor S stars is the oldest among them. 

If the kinematics of Tc-rich M stars may be trusted, it would imply ages of as much as 7~Gyr, which would imply low-metallicity progenitor initial masses as low as 1~M$_\odot$, and forbid progenitors with initial masses 2~M$_\odot$. 
 
\section{Discussion}\label{sec:discussion}
From the previous sections, the following consistent picture seems to emerge concerning Tc-rich M stars:

\begin{itemize}

\item From the abundance analysis (Sect.~\ref{sec:meta} and Table~\ref{tab:meta}), Tc-rich M stars appear to cover a metallicity range $-0.9 \leq $[Fe/H] $\leq -0.3$.
Although it is based only on the three least variable stars for which an abundance analysis could be performed, this hints at Tc-rich M stars having subsolar metallicities.

\item From their predicted Tc and Zr abundances (grey box in Fig.~\ref{fig:nucleopredictions}), Tc-rich M stars are marginally compatible with stellar evolution and nucleosynthesis models corresponding to initial masses 1.5 - 3 \msun\ at metallicity [Fe/H] $= -0.5$.
%$=-1$ metallicities.

\item From their TiO and ZrO band depths
and the $B_{\TiO}$– $B_{\ZrO}$ plane (Fig.~\ref{fig:ZrOTiO}),
the best-fitting synthetic spectra for M stars with the strongest
ZrO and TiO absorption correspond to the lowest metallicity
models ([Fe/H]~$= -1$~dex). This effect may be attributed to $\alpha$-element enrichment, which is anti-correlated with metallicity,
allowing only the lowest-metallicity synthesis to reproduce the highest $B_{\TiO}$ values. However, some Tc-rich M stars are also found along the solar-metallicity tracks.

\item In the HR diagram (Sect.~\ref{sec:HRD}), the location of Tc-rich M stars is compatible with stars of masses $\sim1.5$~\msun, amidst Tc-rich S stars. 

\item  The distribution of Tc-poor and Tc-rich M stars above the Galactic plane is similar (190~pc), and very close as well to that of the `background'\footnote{as opposed to M giants involved in specific stellar streams, as described by \citet{Famaey2005}.} M giants studied by \citet{Famaey2005}.  
Tc-rich M stars appear to have a scale height above the galactic plane intermediate between that of Tc-rich S stars (150~pc) and Tc-poor S stars (260~pc). This conclusion may seem at variance with their respective locations in the HR diagram of Fig.~\ref{fig:HRD}, where the range of masses spanned by Tc-rich S stars overlaps that of Tc-rich M stars, while at the same time being much more extended. We stress however that this is almost certainly the result of the smallness of the samples considered in Fig.~\ref{fig:HRD}, since the scale height obtained from Fig.~\ref{fig:Z_CDF} is derived from a much larger sample less-prone to statistical biases.
\item The velocity dispersions $\sigma_u, \sigma_v, \sigma_w$ of Tc-rich M stars point at a particularly old population, up to 7 - 10~Gyr (Table~\ref{tab:velo}). This is not the case for the other subgroups of M giants, and this higher age therefore seems to be a distinctive property of Tc-rich M giants. The reason for this old age is however not easily identified. Giants with masses 
on the order of 1.0~\msun~match the requirement of 10-Gyr of age.
However, a well-known limitation of current nucleosynthesis models is their failure to trigger TDU at such low masses, at least when using standard prescriptions for mixing processes. Nevertheless, observational evidence for s-process enrichment in stars with initial masses near 1.0~\msun~does exist, such as Tc-rich S stars reported by \cite{shetye2019} and low-mass, s-process-enhanced post-AGB stars identified by \cite{DeSmedt2015} and \cite{Kamath2022}.
The other subgroups of M giants, which do not face the constraint of s-process nucleosynthesis followed by TDUs, could populate the giant branch instead of the AGB. Hence their masses could span the range 1 - 2~\msun, resulting in an average age much smaller than that of Tc-rich M stars. Why AGB stars with masses as high as 2~\msun\ do not end up as Tc-rich M stars, despite their predicted passage through the Tc-rich M locus (grey box in Fig.~\ref{fig:nucleopredictions}), has not however been elucidated.
\item The kinematics confirm the conclusion from the Galactic scale heights that Tc-rich M stars are an older population than Tc-rich S stars.
\end{itemize}

All these elements nicely converge to a scenario where Tc-rich M stars are subsolar metallicity AGB stars with masses as low as 1~\msun,
whereas Tc-rich S stars would be slightly more massive on average, based on their hotter kinematics and smaller Galactic scale height. 

The only clear difference with Tc-rich S stars is their enhanced TiO bands. The reason of these enhanced TiO bands is not currently clear: it might be due to a  larger O abundance 
or a cooler temperature. This latter hypothesis is not confirmed by the HR diagram. However, we cannot exclude these stars to have cool starspots covering a significant fraction of their surface. Since these spots are cooler, they would contribute to stronger TiO features. Such spots could be related to the larger photometric variability of Tc-rich M stars, since the photometric variability increases with increasing TiO and ZrO bands (see Fig.~\ref{fig:ZrOTiO_var}).

Lastly, the classification of S- and M-type stars is traditionally based on ZrO and TiO band strengths, where the M stars are expected to show only TiO bands while the S stars are expected to show both ZrO and TiO bands. 
However, Fig.~\ref{fig:ZrOTiO} shows that Tc-rich M stars also exhibit high $B_{\ZrO}$ and $B_{\TiO}$ indices, with some Tc-poor M stars populating the same region. 
While the elevated $B_{\ZrO}$ in Tc-rich M stars may be due to a slight Zr enhancement (at most [Zr/Fe] $\sim$ 0.2 dex according to Fig.~\ref{fig:ZrOTiO}), the high $B_{\ZrO}$ in some Tc-poor M stars is unexpected.
The Tc-poor M stars with high $B_{\ZrO}$ and $B_{\TiO}$ indices also exhibit strong variability (Fig.~\ref{fig:ZrOTiO_var}). Furthermore, band indices derived from the synthetic spectra (Fig.~\ref{fig:ZrOTiO}) indicate that the $B_{\ZrO}$ and $B_{\TiO}$ plane is highly sensitive to temperature. 
This suggests that the highly variable and cool Tc-rich and Tc-poor M stars could be misclassified as M-type as the relative band strengths of ZrO-TiO are strongly hindered by the temperature effects.

\section{Conclusion}\label{sec:conclusion}
In this study, we performed a detailed analysis of a large sample of \totalM~M-type stars using high-resolution HERMES spectra. 
These stars were discovered thanks to the increasingly large number of available high-resolution spectra, which allow the analysis of the weak blue Tc~I lines, whereas previous spectral classification of M and S stars was merely based on TiO and ZrO bands, well-visible on low-resolution spectra.

We confirmed the Tc-rich nature of \tcrichM~M stars, while \tcpoorM~M stars in our sample were identified as Tc-poor.
We investigated the evolution of Zr and Tc abundances pulse-by-pulse during the thermally pulsing AGB phase in nucleosynthesis models. 
Our models show that a significant Zr enrichment only emerges at \logTc\ values between -2 and 0. 
Furthermore, our spectral synthesis modeling shows that the detection threshold for classifying an M-type star as Tc-rich is \logTc~$=-0.5$.
At \logTc~$= -2$ to 0 (Fig.~\ref{fig:nucleopredictions}), the nucleosynthesis predictions indicate only a mild enhancement in [Zr/Fe] of up to $\sim$ 0.2 dex.
These findings suggest that some M stars can exhibit clear Tc absorption features without significantly strong ZrO bands. 
A comparison of the measured and synthetic spectral indices (Fig.~\ref{fig:ZrOTiO}) demonstrates that the $B_{\ZrO}$ index from synthetic spectra can encompass the regions occupied by Tc-rich M stars, even when the s-process enrichment
is relatively mild, with [s/Fe] between 0 and 0.2 dex.

The spectral indices of M-type stars exhibit several intriguing characteristics: while their Tc features closely resemble those of S-type stars, they follow a distinct sequence in the $B_{\TiO}$ - $B_{\ZrO}$ plane. 
Analysis of these spectral band indices suggests that Tc-rich M stars may have a sligtly lower metallicities than Tc-rich S stars. 
To test this hypothesis, we determined the metallicity of three Tc-rich M stars, finding values between -0.3 and -0.9 (see Table~\ref{tab:meta} and Sect.~\ref{sec:meta}). 
These results indicate that low metallicity, and its influence on s-process nucleosynthesis, could contribute to the presence of Tc in these stars.
Furthermore, by comparing the spatial and kinematical properties of our sample with those of intrinsic and extrinsic S-type stars as well as N-type carbon stars, we find that Tc-rich and Tc-poor M stars belong to the same population, with Tc-rich stars having undergone TDU. 
Tc-rich M stars may also be older than Tc-rich S stars, further supporting a metallicity distribution that is slightly shifted toward lower values compared to Tc-rich S stars. 

Although the origin of the enhanced TiO bands is not fully understood, it likely explains the existence of Tc-rich M stars. 
These stars have been classified as M-type rather than S-type precisely because of their prominent TiO absorption features. 
If their TiO bands were weaker, given that their ZrO band strengths are comparable to those of S-type stars, they might have been classified instead as Tc-rich S stars. Thus, their existence may simply reflect a classification bias.

\begin{acknowledgements}
SS would like to acknowledge the support of Research Foundation-Flanders (grant number: 1239522N). 
AE received the support of a fellowship from ``La Caixa” Foundation (ID 100010434) with fellowship code LCF/BQ/PI23/11970031.
SG and LS acknowledge financial support from F.R.S.-FNRS (Belgium).
SU acknowledges the support from the Austrian Science Fund (10.55776/F81). 
Based on observations made with the Mercator Telescope, operated on the island of La Palma by the Flemish Community, at the Spanish Observatorio del Roque de los Muchachos of the Instituto de Astrofísica de Canarias. 
The IRI number for Mercator operations is I000325N.
This work has made use of data from the European Space Agency (ESA) mission {\it Gaia} (\url{https://www.cosmos.esa.int/gaia}).
For open access purposes, the author has applied a CC BY public copyright license to any author accepted manuscript version arising from this submission.
\end{acknowledgements}

%-------------------------------------------------------------------
\bibliographystyle{aa}
\bibliography{library}

%\end{document}
\begin{appendix}
\section{Variability and Molecular Band Strengths}
\begin{figure}
    \centering
    \includegraphics[width=1.01\linewidth]{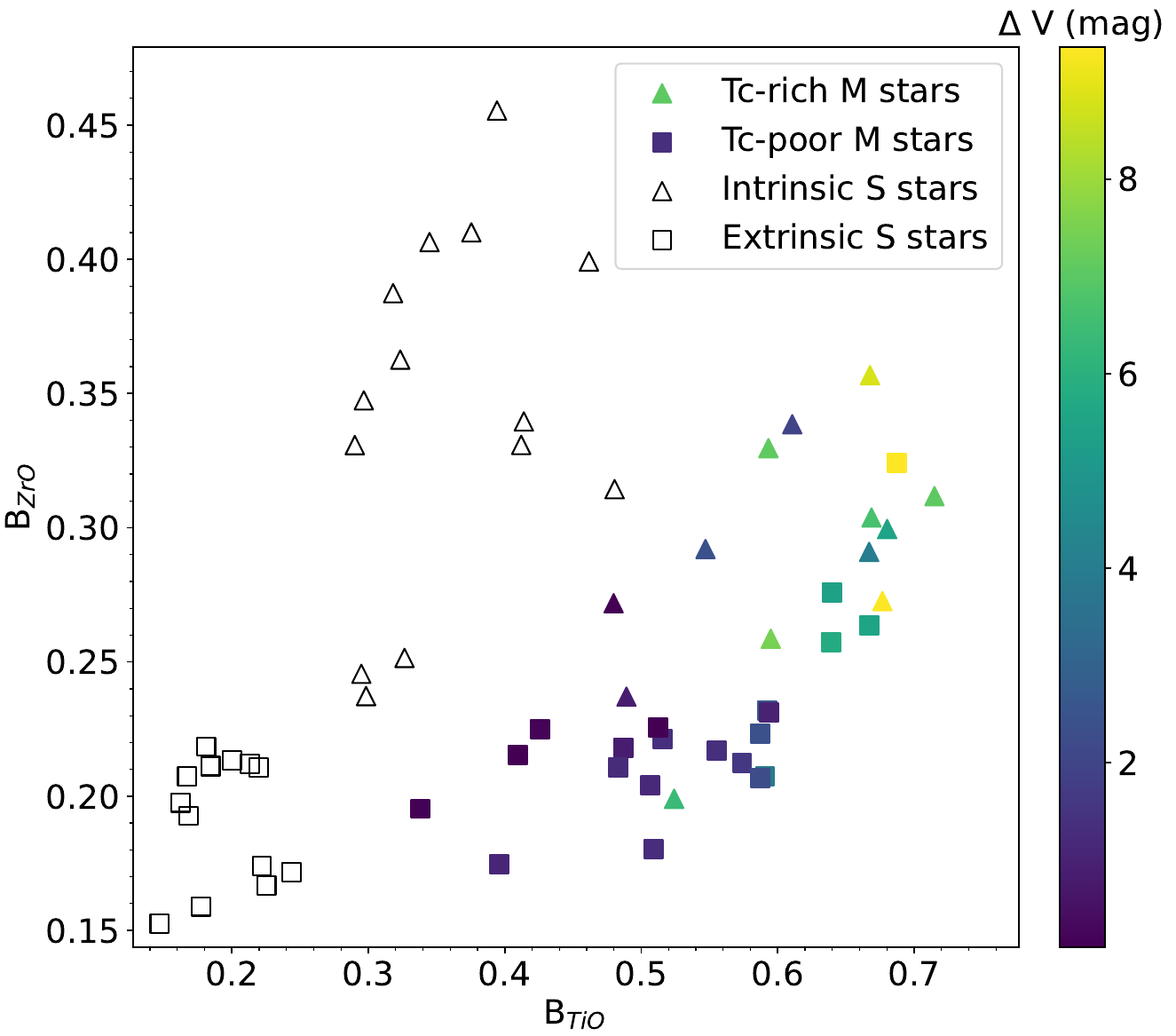}
    \caption{The measured $B_{\TiO}$ and $B_{\ZrO}$, same as Figure~\ref{fig:ZrOTiO}. The sample M stars are color coded with respect to their $\Delta$V listed in Table~\ref{tab:basicdata}. }
    \label{fig:ZrOTiO_var}
\end{figure}

Figure~\ref{fig:ZrOTiO_var} shows the measured $B_{\TiO}$ and $B_{\ZrO}$ indices for the sample M stars, color-coded by their variability amplitude ($\Delta$~V) as listed in Table~\ref{tab:basicdata}. 
This visualization allows for an assessment of whether stellar variability correlates with the strength of TiO and ZrO molecular bands.
Please note that only the variability of our sample M stars have been included here because this information was not available for the majority of the S stars from \cite{abc2021}.

\section{Basic data of the sample stars}
\begin{figure*}[!ht]
    \centering
    \includegraphics[scale=0.55]{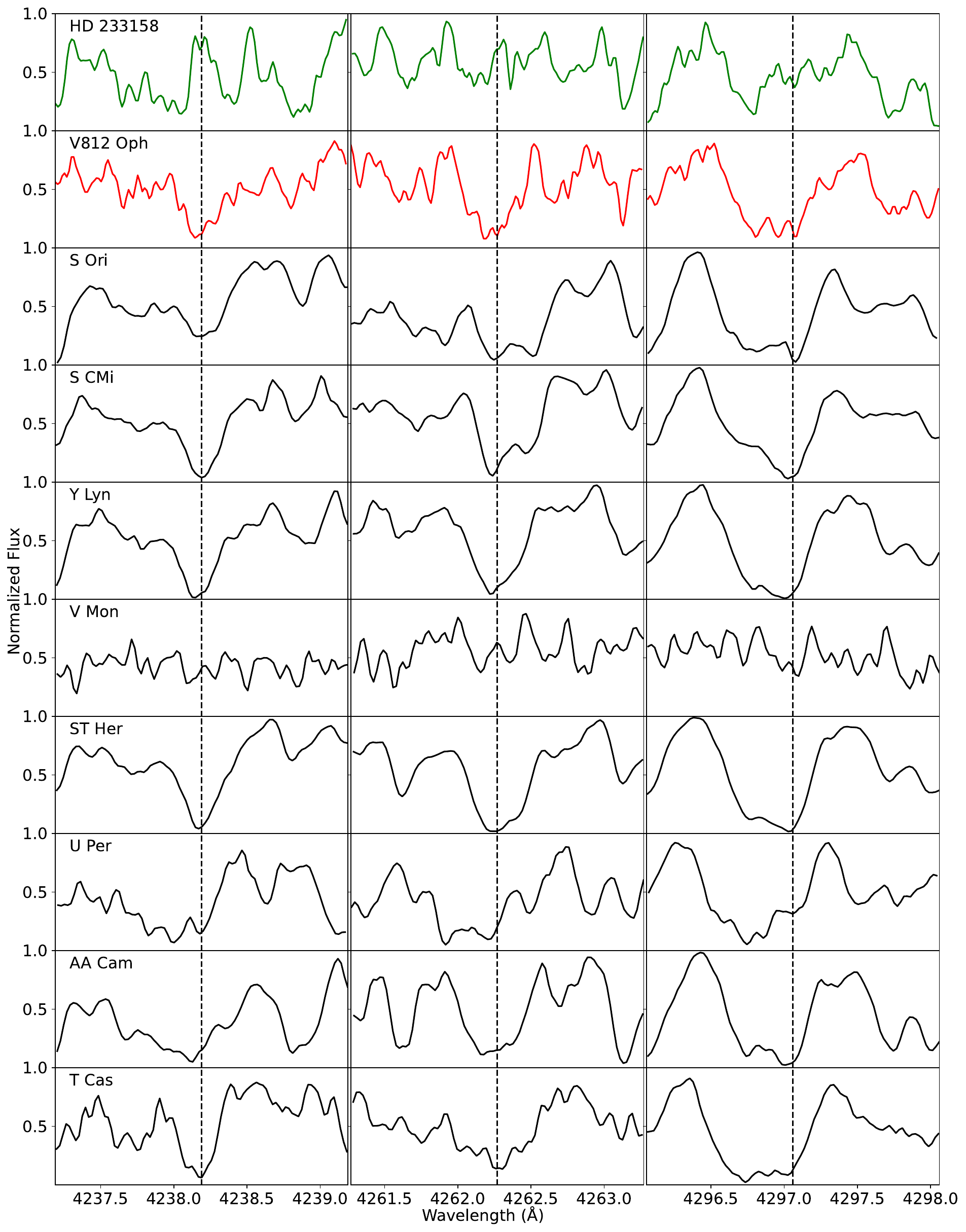}
    \caption{The spectral region around the three (4238.19, 4262.27 and 4297.06 \AA) Tc I resonance lines in our sample of Tc-rich M stars. For comparison purposes, the spectrum of a Tc-poor S star (HD 233158, in green in the top panels) from S18 and a Tc-rich S star (V812 Oph, in red) from S21 are also plotted. \referee{The spectra have been arbitrarily normalized and smoothed using a Gaussian filter with a standard deviation (sigma) of 1\AA~to slightly increase the S/N ratio.} }
    \label{fig:Tcrichset1}
\end{figure*}

\begin{figure*}[!ht]
    \centering
    \includegraphics[scale=0.55]{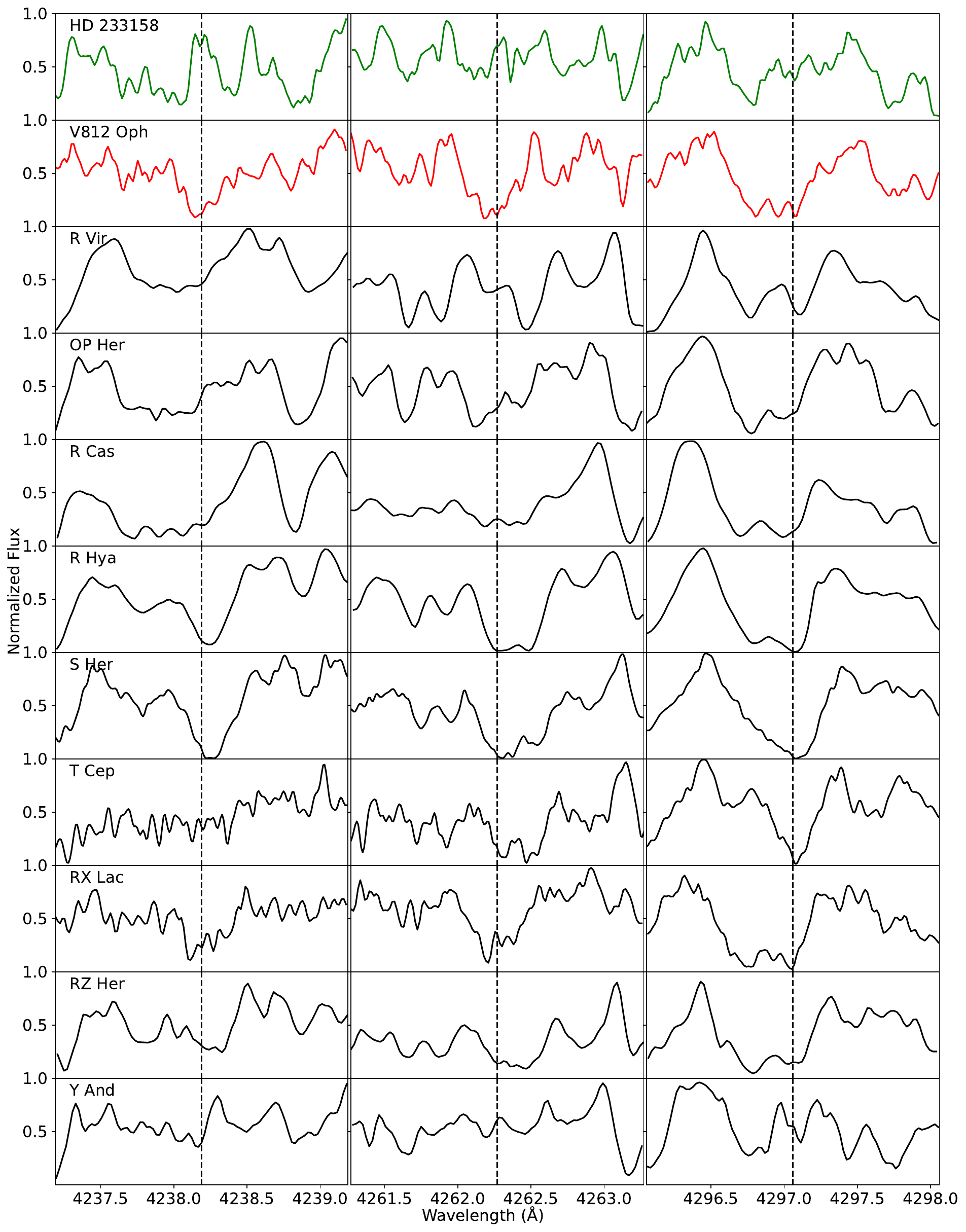}
    \caption{Same as Fig.~\ref{fig:Tcrichset1} for the remaining Tc-rich M stars in our sample.}
    \label{fig:Tcrichset2}
\end{figure*}

\begin{figure*}[!ht]
    \centering
    \includegraphics[scale=0.55]{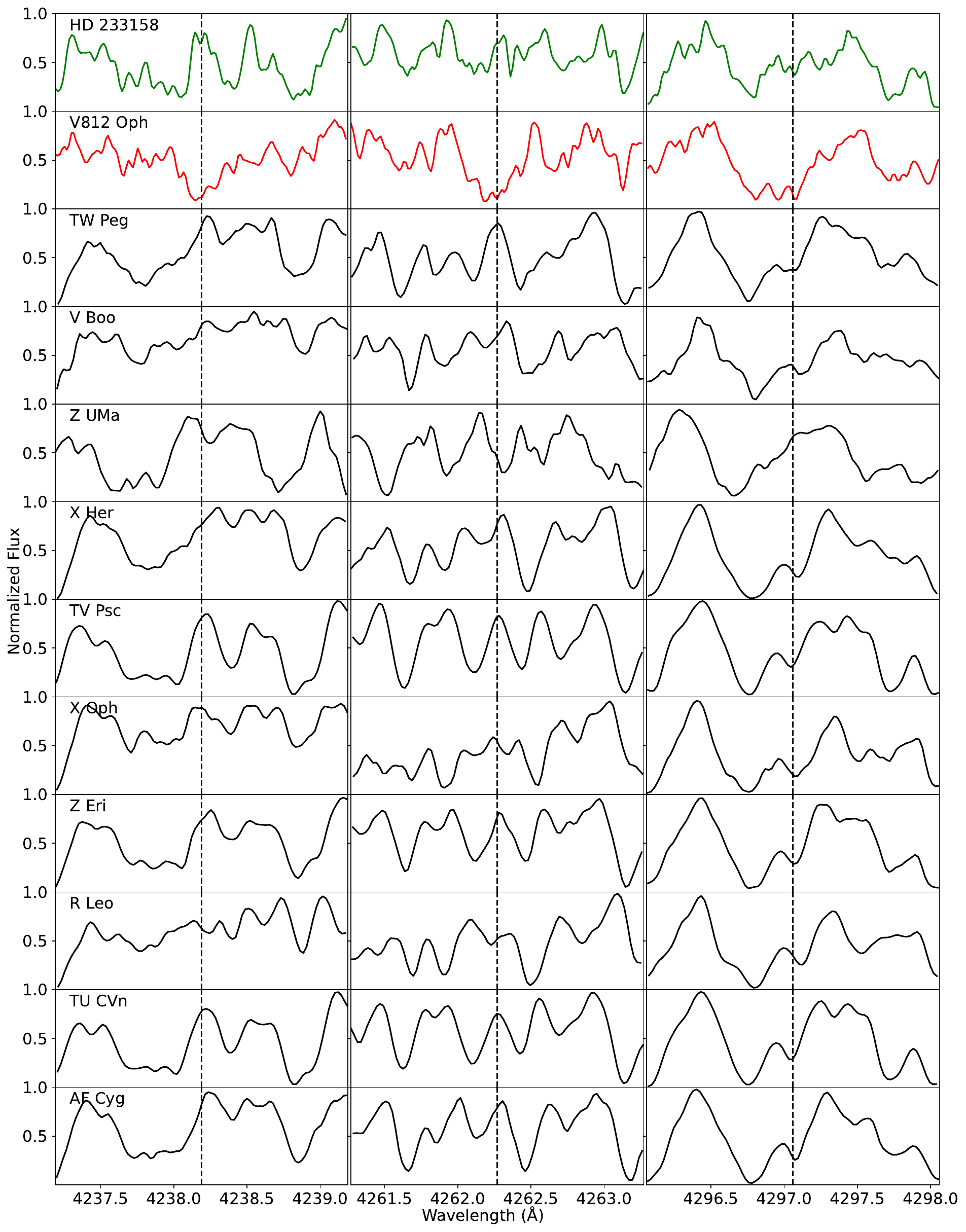}
    \caption{Same as Fig.~\ref{fig:Tcrichset1} for the Tc-poor M stars in our sample.}
    \label{fig:Tcpoorset1}
\end{figure*}

\begin{figure*}[!ht]
    \centering
    \includegraphics[scale=0.55]{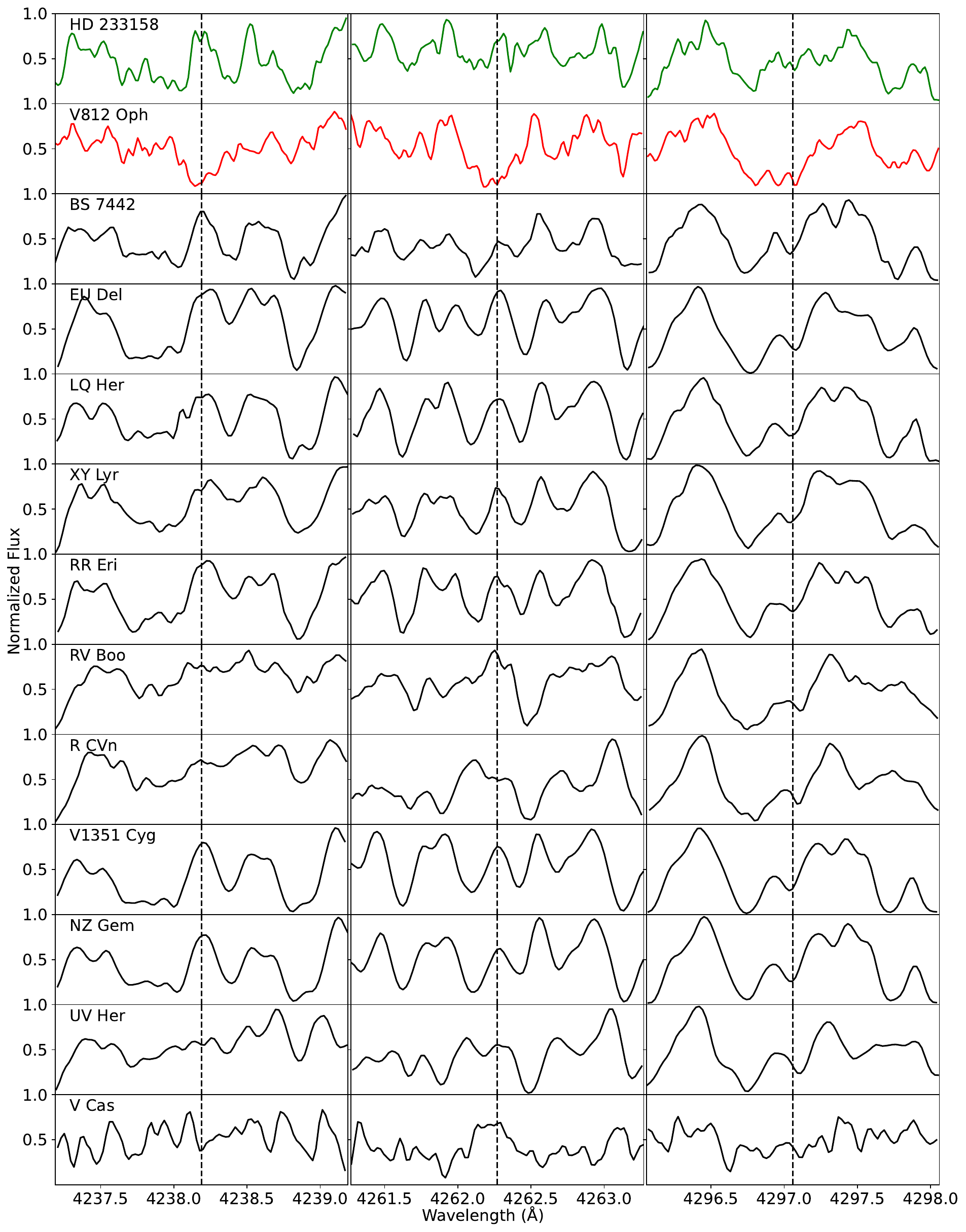}
    \caption{Same as Fig.~\ref{fig:Tcrichset1} for the Tc-poor M stars in our sample.}
    \label{fig:Tcpoorset2}
\end{figure*}

In this section, we provide an overview of the fundamental and spectral properties of the sample stars. 
In Table~\ref{tab:basicdata}, the basic properties are summarized along with the sample classification into Tc-rich and Tc-poor groups.
In Tables~\ref{tab:spectral_indices}, the spectra indices computed in Sect.~\ref{sec:TiOZrO} are enlisted for the M stars sample, while in Table~\ref{tab:spectral_indices_Sstars} the same for the S stars from our previous work are listed.
Finally, in Figs.~\ref{fig:Tcrichset1}, \ref{fig:Tcrichset2}, \ref{fig:Tcpoorset1} and \ref{fig:Tcpoorset2} the Tc features for the sample stars are presented.
\onecolumn
\begin{table*}[]
\caption[]{\label{tab:basicdata} Basic data of our M star sample.}
    \centering
    \begin{tabular}{l|llrclcc}
    \hline
    \hline
        Star & Sp. type & $V$  & $K$  & $\omega \pm \sigma_\omega$ & $\Delta V$ & Variability type & |Z| \\
            &    & (mag) & (mag) & (mas) & (mag) & & (pc) \\
            \hline
        \multicolumn{8}{c}{Tc-rich M stars }\\ \hline
        S Ori &  M6.5-7.5e & 7.20 & -0.15 & 2.39 $\pm $ 0.10 & 6.43 & Mira & 146 \\ 
        S CMi &  M7-8e & 6.60 & 0.56 & 2.39  $\pm $ 0.09 & 6.07 & Mira & 93  \\ 
        Y Lyn &  M6S & 6.98 & -0.44 & 2.80  $\pm $ 0.11 & 2.27 & SRV & 152  \\ 
        V Mon &  M5-8Se & 6.87 & 1.17 & 1.40  $\pm $ 0.09 & 7.94& Mira & 91   \\ 
        ST Her &  M6S-M7SIII: & 7.75 & -0.54 & 3.39  $\pm $ 0.11 & 1.88 & SRV &224  \\ 
        U Per &  M5.5-7e & 8.76 & 0.86 & 2.10$\pm $ 0.08 & 5.89  & Mira &56 \\ 
        AA Cam &  M5S & 7.79 & 1.39 & 2.08$ \pm $ 0.05 & 0.24 & SRV &220  \\ 
        T Cas &  M7-9e & 8.87 & -0.78 & 3.63   $ \pm $  0.17 & 3.69 & Mira &33  \\ 
        R Vir &  M3.5-7e & 7.48 & 1.99 & 2.19   $ \pm $  0.05 & 5.81& Mira &427  \\ 
        OP Her &  M6S & 6.32 & 0.07 & 3.45   $ \pm $ 0.06 & 0.90 & SRV & 136   \\ 
        R Cas &  M6.5-9e & 4.80 & -1.92 & 5.74   $ \pm $ 0.20 & 8.46 & Mira &32   \\ 
        R Hya &  M6-9e & 4.97 & -2.51 & 6.73   $ \pm $ 0.46 & 5.05 & Mira &93  \\ 
        S Her & M4-7.5Se & 6.40 & 1.25 & 1.71   $ \pm $  0.05 & 6.44 & Mira &319  \\ 
        T Cep &  M6-9e & 7.33 & -1.82 & 5.44   $ \pm $ 0.34 & 4.56 & Mira &44   \\ 
        R Aqr &  M6.5-8.5e & 7.68 & -1.60 & 2.59 $ \pm $  0.33 & 6.81 & Mira &363 \\ 
        RX Lac &  S7.5/1e & 8.62 & -0.02 & 2.91  $ \pm $ 0.15 & 1.54 & SRV &96  \\ 
        RZ Her &  M5e & 11.27 & 3.96 & 0.52  $ \pm $ 0.04 & 6.77 & Mira &478  \\
        Y And &  M3-4.5e & 10.39 & 4.15 & 0.55   $ \pm $ 0.06 & 7.12 & Mira & 689 \\ \hline
         \multicolumn{8}{c}{Tc-poor M stars }\\ \hline 
        TW Peg & M6-7III & 7.50 & -0.63 & 4.04$ \pm $  0.13 & 1.33 & SRV &91   \\ 
        V Boo &  M5+-6.5e & 8.46 & 1.035 & 2.09  $ \pm $ 0.05 & 2.60 & SRV &437 \\ 
        Z UMa &  M5IIIv & 6.20 & 0.52 & 3.38  $ \pm $   0.07 & 3.80& Mira & 250   \\ 
        X Her &  M6III & 6.58 & -1.2 & 8.16   $ \pm $    0.25 & 1.60 & SRV & 91  \\
        TV Psc &  M3III & 5.06 & -0.13 & 6.00   $ \pm $   0.25 & 1.10 & SRV &117\\ 
        X Oph &  M0-8e+K2:III & 6.40 & -0.79 & 5.14   $ \pm $   0.34 & 2.41& Mira &23   \\ 
        Z Eri &  M5III & 6.90 & 0.34 & 3.04   $ \pm $   0.10 & 1.16& SRV &281 \\ 
        R Leo &  M7-9e & 7.53 & -2.51 & 14.05$^a$ $ \pm $ 0.83 & 5.36& Mira & - \\ 
        TU CVn &  M5-III-IIIa & 5.84 & -0.2 & 4.17  $ \pm $    0.16 & 0.85& SRV &225   \\ 
        AF Cyg &  M5III & 7.18 & 0.44 & 3.73  $ \pm $   0.12 & 2.25& SRV &60   \\ 
        BS 7442 &  M4.5III & 5.96 & 0.51 & 2.74  $ \pm $   0.07 & 0.10 & SRV &87 \\ 
        EU Del &  M6III & 5.89 & -1.11 & 8.43  $ \pm $   0.28 & 1.27& SRV &28   \\ 
        LQ Her & M3III & 5.70 & 0.05 & 4.74   $ \pm $  0.13 & 0.20 & SRV & 149\\ 
        XY Lyr & M4.5-M5+II & 6.05 & -0.3 & 2.39  $ \pm $   0.09 & 1.24& SRV &138   \\ 
        RR Eri & M5III & 7.16 & 0.42 & 2.63  $ \pm $   0.12 & 1.33 & SRV &314  \\ 
        RV Boo & M5-6IIIe & 7.9 & -0.05 & 2.71  $ \pm $   0.12 & 1.00 & SRV &337  \\ 
        R CVn &  M6.5-9e & 7.40 & 0.5 & 2.17  $ \pm $  0.12 & 5.75 & Mira &440   \\ 
        V1351 Cyg & M5III & 6.48 & 0.34 & 3.36 $ \pm $   0.05 & 0.10 & SRV & 79   \\
        NZ Gem &  M3+II-III & 5.56 & 0.6 & 2.72   $ \pm $  0.12 & 0.15 & SRV & 110\\ 
        UV Her & M6e & 10.38 & 1.86 & 0.79   $ \pm $  0.07 & 5.50 & Mira & 692 \\ 
        V Cas &  M5-7.5e & 6.9 & 0.75 & 2.51 $\pm$ 0.06 & 9.36 & Mira & 5 \\ 
        \hline 
        % \multicolumn{8}{c}{Tc-doubtful M stars }\\ \hline 
        %RT  Cyg &  M2-7e & 8.45 & 3.13 & 0.96   $ \pm $  0.04 & 6.11& LPV &   \\ 
        \hline
    \end{tabular}
    \tablefoot{ Column 2 provides the spectral type, while Columns 3 and 4 show the $V-$and $K-$magnitudes, respectively, as obtained from the SIMBAD\footnote{\url{https://simbad.cds.unistra.fr/simbad/}} Astronomical Database. Column 5 contains the parallax values along with their associated errors, sourced from the \textit{Gaia} Data Release 3. Column 6 reports $\Delta V$, which represents the amplitude variation of the $V-$magnitude, collected from the AAVSO\footnote{\url{https://www.aavso.org/}} database. In the seventh column we present the variability type of our sample stars based on the light curves from the KWS survey \citep{Maehara2014}, where `SRV' stands for semi-regular variable and `Mira' for Mira-type variables. \referee{In the last column we enlist the height above the Galactic plane |Z|.}\\
    $^a$ \textit{Gaia} data release 2 parallax}
    %and `ECL' for eclipsing binaries of type $\beta$ Persei (Algol). }
\end{table*}

\twocolumn
\begin{table}
\centering
\sisetup{round-mode=places}
\setlength{\tabcolsep}{2pt}
 \caption{\label{tab:spectral_indices} Spectral indices of the M stars sample.}
 \begin{tabular}{l|S[round-precision=3]S[round-precision=3]S[round-precision=2]S[round-precision=2]}
    \hline
    \hline
        Star & $\lambda_{4238}$ & $\lambda_{4262}$ & $B_{\TiO}$ & $B_{\ZrO}$ \\ \hline
        \multicolumn{5}{c}{Tc-rich M Stars} \\ \hline
        S Ori & 4238.22 & 4262.2516 & 0.714870044 & 0.311797 \\ 
        S CMi & 4238.207 & 4262.371 & 0.668719993 & 0.3038035505 \\ 
        Y Lyn & 4238.162 & 4262.2473 & 0.54708634 & 0.2920653 \\ 
        V Mon & 4238.2428 & 4262.2322 & 0.667657296 & 0.35682944 \\ 
        ST Her & 4238.0758 & 4262.3909 & 0.6106825 & 0.33855144 \\ 
        AA Cam & 4238.0883 & 4262.2418 & 0.47977 & 0.271871 \\
        T Cas & 4238.1637 & 4262.2561 & 0.6670067322 & 0.2909958 \\ 
        R Vir & 4238.0277 & 4262.279 & 0.5240924 & 0.1990222 \\ 
        OP Her & 4237.9615 & 4262.2335 & 0.48918 & 0.23708 \\
        R Cas & 4238.0571 & 4262.286 & 0.676781 & 0.272658 \\ 
        R Hya & 4238.2441 & 4262.3821 & 0.6802043962 & 0.299531197 \\ 
        S Her & 4238.2564 & 4262.3 & 0.593216168 & 0.329646022 \\ 
        RZ Her & 4238.1 & 4262.328 & 0.5949279 & 0.25864207 \\ \hline
        \multicolumn{5}{c}{Tc-poor M Stars} \\ \hline
        TW Peg & 4238.3756 & 4262.0763 & 0.5555404086 & 0.2169212683 \\ 
        V Boo & 4238.369 & 4262.1028 & 0.5924136229 & 0.2319623962 \\ 
        Z UMa & 4238.4243 & 4262.1371 & 0.590753104 & 0.2075152091 \\ 
        X Her & 4238.3893 & 4262.096 & 0.5738139768 & 0.2124982031 \\ 
        TV Psc & 4238.3869 & 4262.1115 & 0.3959886777 & 0.1746332018 \\ 
        X Oph & 4238.3544 & 4262.1604 & 0.5871635701 & 0.223200524 \\ 
        Z Eri & 4238.394 & 4262.1302 & 0.506643005 & 0.203953049 \\ 
        R Leo & 4238.3409 & 4262.0984 & 0.639737209 & 0.275931776 \\ 
        TU CVn & 4238.3891 & 4262.1027 & 0.487236628 & 0.218111167 \\ 
        AF Cyg & 4238.3866 & 4262.1452 & 0.5874166337 & 0.206739915 \\ 
        BS 7442 & 4238.3818 & 4262.1466 & 0.409696927 & 0.215404135 \\ 
        EU Del & 4238.3844 & 4262.1144 & 0.50885393 & 0.18021542 \\ 
        LQ Her & 4238.3772 & 4262.0982 & 0.42598871 & 0.224889588 \\ 
        XY Lyr & 4238.4029 & 4262.111 & 0.4828900073 & 0.210631281 \\ 
        RR Eri & 4238.3996 & 4262.1159 & 0.51543026 & 0.221337188 \\ 
        RV Boo & 4238.3112 & 4261.9938 & 0.5938085 & 0.231103 \\ 
        R CVn & 4238.4038 & 4261.9544 & 0.63933163 & 0.257284736 \\ 
        V1351 Cyg & 4238.3792 & 4262.0988 & 0.51212324 & 0.22564523 \\ 
        NZ Gem & 4238.3881 & 4262.1104 & 0.33834036 & 0.1953084066 \\ 
        UV Her & 4238.3525 & 4262.0941 & 0.667122 & 0.26365063 \\ 
        V Cas & 4238.404 & 4262.1225 & 0.6871382 & 0.324166 \\ \hline
        \hline
    \end{tabular}
\end{table}

\begin{table}
\centering
\sisetup{round-mode=places}
\setlength{\tabcolsep}{2pt}
 \caption{\label{tab:spectral_indices_Sstars} Spectral indices of the S stars from \cite{shetye2018} and \cite{abc2021}.}
 \begin{tabular}{l|S[round-precision=3]S[round-precision=3]S[round-precision=2]S[round-precision=2]}
    \hline
    \hline
        Star & $\lambda_{4238}$ & $\lambda_{4262}$ & $B_{\TiO}$ & $B_{\ZrO}$ \\ \hline
        \multicolumn{5}{c}{Tc-rich S Stars} \\ \hline
        CSS 89 & 4238.1392 & 4262.2353 & 0.4139517 & 0.33961 \\ 
        CSS 106 & 4238.2355 & 4262.2005 & 0.2901092484 & 0.330798 \\ 
        CSS 114 & 4238.1025 & 4262.2298 & 0.326538 & 0.2515149 \\ 
        CSS 151 & 4238.057 & 4262.2246 & 0.323402021 & 0.36264388 \\ 
        CSS 154 & 4238.0606 & 4262.2299 & 0.46331246 & 0.320806398 \\ 
        CSS 182 & 4238.157 & 4262.2161 & 0.3448760985 & 0.4064166 \\ 
        CSS 233 & 4238.1131 & 4262.234 & 0.298353 & 0.2372964 \\ 
        CSS 265 & 4238.1485 & 4262.2383 & 0.480474339 & 0.31438045 \\ 
        CSS 312 & 4238.1013 & 4262.2244 & 0.2948841555 & 0.2455793 \\ 
        CSS 413 & 4238.1329 & 4262.2566 & 0.4616573391 & 0.39919538 \\ 
        CSS 416 & 4238.096 & 4262.2376 & 0.41202417 & 0.33085902 \\ 
        CSS 454 & 4238.1621 & 4262.2543 & 0.3755039 & 0.410036708 \\ 
        CSS 489 & 4238.1346 & 4262.205 & 0.29669416 & 0.347455 \\ 
        CSS 597 & 4238.1935 & 4262.2382 & 0.3181587 & 0.387404 \\ 
        CSS 997 & 4238.1915 & 4262.2314 & 0.39430309 & 0.455438 \\ 
        CSS 1070 & 4238.2366 & 4262.2847 & 0.0523934 & 0.5627953 \\ \hline
        \multicolumn{5}{c}{Tc-poor S Stars} \\ \hline
        AB Col & 4238.3872 & 4262.1033 & 0.21971574 & 0.2108056 \\ 
        BD -22 1742 & 4238.3645 & 4262.0978 & 0.1685181 & 0.19268274 \\ 
        BD +69 524 & 4238.3798 & 4262.122 & 0.200153 & 0.213244 \\ 
        BD 284592 & 4238.3846 & 4262.0914 & 0.1622699 & 0.197612 \\ 
        FX CMa & 4238.3488 & 4262.0996 & 0.184302 & 0.211266 \\ 
        HD 150922 & 4238.3837 & 4262.1033 & 0.1771789 & 0.1589061 \\ 
        HD 189581 & 4238.3928 & 4262.1121 & 0.2436218 & 0.171593 \\ 
        HD 191226 & 4238.3808 & 4262.0995 & 0.2132848 & 0.21204734 \\ 
        HD 191589 & 4238.3802 & 4262.0923 & 0.1471063 & 0.1526027 \\ 
        HD 215336 & 4238.3789 & 4262.0893 & 0.22193465 & 0.17399364 \\ 
        HD 233158 & 4238.3718 & 4262.1345 & 0.225264 & 0.166739 \\ 
        HIP 38217 & 4238.3842 & 4262.1029 & 0.181021 & 0.2184628 \\ 
        V0530 Lyr & 4238.3827 & 4262.0979 & 0.18430599 & 0.210907 \\ 
        V1135 Tau & 4238.3909 & 4262.139 & 0.1672864 & 0.2072796 \\ \hline
        \hline
    \end{tabular}
\end{table}

%\section{Technetium lines of the sample stars}

\end{appendix}

\end{document}